\documentclass[12pt]{article}

\usepackage{geometry}
\usepackage[utf8]{inputenc}

\usepackage{amsmath,amssymb,amsthm,bbm}

\usepackage{graphicx}
\graphicspath{{figures-cooperation/}}

\usepackage[table]{xcolor}
\usepackage{booktabs,tabularx,threeparttable,longtable,array,multirow,dcolumn}

\usepackage{float}
\usepackage{caption}
\usepackage{subcaption}

\usepackage[comma,authoryear]{natbib}

\usepackage{setspace}
\usepackage{soul}
\usepackage{pdflscape}
\usepackage{placeins}
\usepackage{todonotes}
\usepackage{blindtext}
\usepackage{csquotes}
\MakeOuterQuote{"}
\usepackage{pgffor}          
\usepackage{pdfcomment}      

\usepackage{hyperref}         
\newtheorem{hypothesis}{Hypothesis}
\newtheorem{result}{Result}

\def\sym#1{\ifmmode^{#1}\else\(^{#1}\)\fi}

\usepackage{titling}
\title{Delegate Pricing Decisions to an Algorithm? \\ Experimental Evidence}

\author{Hans-Theo Normann\thanks{normann@hhu.de. Düsseldorf Institute for Competition Economics (DICE) at Heinrich-Heine-Universität Düsseldorf, Max Planck Institute for Research on Collective Goods, Bonn} \and Nina Rulié\thanks{nina.rulie@just.fgov.be. Service d’appui du Collège des cours et tribunaux / Steundienst van het College van hoven en rechtbanken, Belgium}
\and Olaf Stypa\thanks{stypa@dice.hhu.de. Düsseldorf Institute for Competition Economics (DICE) at Heinrich-Heine-Universität Düsseldorf}\and Tobias Werner\thanks{tobias.felix.werner@soton.ac.uk. University of Southampton}}

\begin{document}

\date{\today}
\maketitle

\renewcommand{\thefootnote}{\fnsymbol{footnote}}
\setcounter{footnote}{5} 
\footnotetext{We are grateful to Zach Brown, Emilio Calvano, and Ed Hopkins for valuable comments. Seminar participants and conference audiences at the following venues gave useful feedback on the
paper: Aarhus University, BSE Summer Forum Workshop on Computational and Experimental Economics (Barcelona), LEE Workshop on Human-AI Interaction, University of Michigan, VfS Annual Conference 2025 (Cologne). We thank Maximilian Andres, Juri Nithammer, and Lisa Bruttel for their support in running sessions in Potsdam.
The pre-registration can be found here https://osf.io/fh89u/. Ethical approval was granted by the German Association for Experimental Economic Research (no. Cmvrj3oD).}
\renewcommand{\thefootnote}{\arabic{footnote}} 
\setcounter{footnote}{0}

\begin{abstract}
\noindent We analyze the delegation of pricing by participants, representing firms, to a collusive, self-learning algorithm in a repeated Bertrand experiment. In the baseline treatment, participants set prices themselves. In the other treatments, participants can either delegate pricing to the algorithm at the beginning of each supergame or receive algorithmic recommendations that they can override. Participants delegate more when they can override the algorithm's decisions. In both algorithmic treatments, prices are lower than in the baseline. Our results indicate that while self-learning pricing algorithms can be collusive, they can foster competition rather than collusion with humans-in-the-loop.

\end{abstract}


\bigskip


\pagebreak
\onehalfspacing

\section{Introduction}
\label{sec:introduction}

The increasing use of pricing algorithms across different markets is sparking significant antitrust concerns. Pricing algorithms are used by traditional businesses, such as gas stations \citep{assad2020algorithmic} and supermarkets \citep{Competera}, and are particularly relevant in the growing e-commerce sector \citep{Chen2016, EuropeanCommission2017FinalInquiry}. This growing reliance on algorithms has raised concerns among regulators and legal scholars about potential anti-competitive effects. Algorithms may tacitly learn to collude without explicit instruction or enable hub-and-spoke collusion when supplied by a common provider \citep[e.g.,][]{BundeskartellamtandAutoritedelaconcurrence2019AlgorithmsCompetition, CompetitionMarketsAuthority2021Algorithms, EzrachiStucke2020}. These concerns are supported by empirical and theoretical work showing that algorithm adoption can facilitate coordination and raise prices \citep[see, for instance,][]{brown2020competition, banchio2022artificial, assad2020algorithmic, musolff2024algorithmic}. Furthermore, simulation studies demonstrate that reinforcement learning algorithms can autonomously learn to sustain supra-competitive pricing strategies \citep{calvano2020artificial, klein2021autonomous}.

This paper considers the decision-making process of individuals when adopting such pricing algorithms. In contrast to previous studies, we do not treat algorithm adoption as an exogenous experimental condition, but as a strategic choice made by humans, such as pricing managers. Humans can often be reluctant to rely on algorithms, even when they are superior. This well-documented phenomenon, known as ``algorithm aversion'' may inhibit the adoption of pricing algorithms \citep{dietvorst_algorithm_2015, Dietvorst2016overcoming}. Moreover, in a repeated pricing setting, the decision not to adopt an algorithm could itself be a profitable deviation, which may in turn provide an incentive to price manually and not use a possibly collusive algorithm. The role of human adoption in shaping the effects of pricing algorithms has received limited attention, and overlooking this dimension may lead to overestimating the risk of algorithmic collusion. Our study addresses this gap by explicitly modeling adoption as a strategic choice and examining how it affects market outcomes.

We conduct laboratory experiments to answer two research questions: First, are participants more likely to adopt a pricing algorithm when they retain the ability to override its recommendations? Second, how does the option to use a price algorithm affect market prices when adoption is left to human decision-makers? 
The algorithm used in our experiment has been trained through extensive simulations with popular reinforcement learning algorithms, similar to those studied by \citet{calvano2020artificial} and \citet{klein2021autonomous}. It learns a collusive strategy and is capable of sustaining tacit collusion when matched against itself or other reinforcement learning algorithms. In algorithmic self-play, it colludes at the monopoly price and cannot be profitably exploited. Participants are made aware of these potential profit-increasing effects of the algorithm throughout the experiment.\footnote{Our setup reflects a scenario where firms knowingly outsource pricing to an algorithm capable of sustaining collusion. From a policy perspective, this may be viewed as a lower bound on the anti-competitive risk of algorithmic pricing, since one would expect adoption rates and potential harms to be higher when firms believe such algorithms are collusive and profitable.} In our laboratory experiment, participants play five indefinitely repeated supergames. We vary whether participants can adopt an algorithm at the start of each of these supergames and how much control they must relinquish to it if they adopt it. In the \textsc{Baseline} treatment, participants set prices themselves. In the \textsc{Recommendation} treatment, participants can adopt an algorithm that suggests prices, but its advice can be overridden. In the \textsc{Outsourcing} treatment, adoption entails full delegation and participants hand over all pricing decisions to the algorithm for the duration of the supergame.\footnote{In practice, both full delegation and algorithmic recommendations are commonly used. For instance, \citet{huelden2024human} and \citet{li2021large} report on a pricing system at a large e-commerce firm in which algorithms are fully responsible for pricing decisions by default, while humans can only intervene to overwrite them. On the other hand, for instance \citet{garcia2024strategic} describe algorithms for hotel room pricing that only serve as recommendations to human pricing managers. As such, both delegation treatments are also relevant from a practical perspective.} Our design allows us to analyze how the anti-competitive effects of pricing algorithms depend on endogenous adoption choices and the degree of control participants retain. It also enables us to investigate algorithm aversion in a pricing context and test whether participants are more inclined to adopt an algorithm when they can override its recommendations, as discussed by \citet{Dietvorst2016overcoming} for non-strategic environments.

Our results show that adoption of the algorithm is substantial. Across treatments and supergames, adoption rates range from 45\% to 86\%. Consistent with prior findings in non-strategic environments on algorithm aversion, we find that adoption is significantly higher in the \textsc{Recommendation} treatment than in \textsc{Outsourcing}, suggesting that participants are more willing to rely on the algorithm when they retain the ability to override its suggestions. However, adoption declines over time in both treatments, indicating that participants reassess the value of using the algorithm as they gain experience and learn from repeated market interactions.

We find only limited evidence that pricing algorithms are pro-collusive with endogenous adoption. While these algorithms are collusive among themselves, this pattern does not carry over when adoption is left to human participants. While market prices are higher in \textsc{Outsourcing} than in the other treatments in the initial supergame, prices in both algorithmic treatments decline in later supergames. We identify repeated override decisions and strategic undercutting of algorithmic prices as key channels that drive those trends and, in turn, discourage further adoption. In the final supergame, market prices in the \textsc{Baseline} treatment, where no algorithm is available, are significantly higher than in both \textsc{Outsourcing} and \textsc{Recommendation}. This reversal highlights the importance of endogenizing adoption. Algorithms that foster collusion when used by all firms can even lead to more competitive outcomes when adoption is selective and humans retain control.

These results highlight the importance of treating algorithm adoption as an endogenous decision influenced by both strategic and behavioral factors, which the existing literature has largely overlooked. Empirical studies show that algorithmic pricing can raise prices, increase markups, or support coordination in markets such as fuel retailing, e-commerce, and rental housing \citep[e.g.,][]{assad2020algorithmic, calder2024algorithmic, musolff2024algorithmic}. Simulation studies show that reinforcement learning algorithms can learn collusive strategies even without being explicitly programmed to do so \citep{waltman2008q, klein2021autonomous, johnson2020platform, martin2025spoils}, and similar results hold for more advanced algorithms that use neural networks or large language models \citep{deng2024algorithmic, hettich2021algorithmic, fish2024algorithmic}. 
However, most of this work assumes that firms adopt algorithms automatically and does not consider the human decision to use them. Theoretical models also often ignore this human element, with a few exceptions that include managerial oversight \citep[e.g.,][]{leisten2024algorithmic}. A few theoretical contributions have begun to formalize related issues, focusing on algorithmic commitment rather than adoption behavior \citep[e.g.,][]{brown2020competition,brown2025algorithmic,leisten2024algorithmic}. Our findings instead highlight how human oversight and delegation choices affect pricing outcomes directly and show that once human adoption and control is taken into account, the potential for algorithmic pricing to raise market prices can be significantly reduced, and in some cases, even reversed.

Our experiment lets participants choose whether to use a pricing algorithm and whether they keep control over its decisions. It allows us to study how algorithm aversion works in a strategic market setting, where choices affect both profits and market outcomes \citep{dietvorst_algorithm_2015, Dietvorst2016overcoming, ivanova2024measuring}. It also enables us to observe how the ability to override the algorithm shapes adoption. Prior research has shown that pricing algorithms can raise prices in markets with human participants \citep{hunold2023algorithmic, normann2021hybrid, schauer2023competition, werner2021algorithmic}, but in those studies, algorithm use was fixed by design. In our study, participants decide for themselves whether and how to use the algorithm.

The remainder of the paper is structured as follows. Section \ref{sec:design} outlines the experimental design, including the treatment conditions, the algorithm, and the hypothesis. Section \ref{sec:results} presents the main results, and Section \ref{sec:discussion} explores the mechanisms that help explain these findings. Section \ref{sec:conclusion} concludes.

\section{Experimental design}\label{sec:design}
\subsection{Market game and equilibria}

The market consists of two firms producing a homogeneous good facing perfectly inelastic demand.\footnote{This Bertrand market experiment was first proposed by \cite{dufwenberg2000price} in a static environment and is commonly used in collusion experiments \citep[see, for example,][]{fonseca2012explicit}. }  
They have no (marginal) cost of production. The firms' action sets are the integer prices \{0, 1, 2, ..., 5\}. Total demand is given by \(m = 60\) computerized consumers who are willing to purchase one unit of the good in each round as long as the price does not exceed the maximum willingness to pay, \(\bar{p} = 4\). Consumers buy the good from the firm offering the lowest price. The market is divided equally if both firms offer the same price in a given period. Depending on the treatment, a human or an algorithm may set the price. This market environment is the same across all experimental treatments. 

Subjects play five infinitely repeated supergames using the above Bertrand duo\-poly as the stage game. In each period, there is a 95\% probability that the supergame will continue for another round, mimicking the infinite time horizons and discount factors inherent in repeated-game analyses as first implemented by \cite{Roth-Murnighan-1978}. This information is known to the participants. Compared to other experiments \citep[see the meta study of][]{dal-bo-frechette-JEL-2018}, the continuation probability is relatively high. This was done to maintain comparability to studies that investigate the collusive potential of self-learning algorithm where simulations are typically conducted with a continuation probability of at least 0.9. In order to have experimental sessions with the same supergame lengths, the round numbers are pre-drawn with a random number generator and are not provided to participants. 

The equilibria of the market game are as follows. In the one-shot game, the set of pure-strategy Nash equilibrium prices is $\{0, 1, 2\}$. 
In the repeated game, more subgame perfect equilibria can occur. Next to the static Nash equilibria, both firms charging a price of 3 or 4 are subgame perfect equilibria with grim trigger strategies. 
The collusive price of 4 maximizes the joint profits: When both firms charge this price, the profit is \(4 \times 30 =120\) each. The deviation profit is \(3 \times 60=180\) and the Nash threat profits for \(p^{NE} = 0\) (the most severe credible punishment) are zero. Adhering to the collusive price of 4 is better than deviating if and only if $120/(1-\delta)\geq 180  \Leftrightarrow \delta \geq 1/3$. Similar calculations for the collusive price of 3 show that adhering is better than deviating if and only if $\delta \geq 1/4$. 
Since our algorithm colludes at a price of 4 but only triggers a one-period punishment (see below), we must also demonstrate that this strategy is a subgame-perfect Nash equilibrium. This is the case if and only if it is not profitable to deviate in the current period and jointly play the stage game Nash equilibrium for one period afterwards. Formally, this condition is fulfilled if and only if $120 + \delta 120 \geq 180 + \delta 30 \Leftrightarrow \delta \geq 6/9$. 
    
We conclude that the prices 0, 1 and 2 can be interpreted as (Nash-)competitive price levels. Prices 3 and 4 are supra-competitive and may indicate tacit collusion. A price of 5 would imply zero demand, and thus probably indicates an error. The termination probability in the experiment is sufficiently high for tacit collusion to be a subgame perfect equilibrium.

\subsection{Treatments}

We consider three experimental treatments in which we modify whether and how participants can use the help of an algorithmic pricing agent. We call these treatments \textsc{Baseline},  \textsc{Outsourcing} and \textsc{Recommendation}. 

In the \textsc{Baseline} treatment, participants do not have the option of using a pricing algorithm in any way. They set the prices themselves, as in a standard Bertrand experiment. 

In the \textsc{Outsourcing} treatment, participants can delegate the pricing decision to the algorithm or play the supergame themselves. This decision is made once, at the beginning of each supergame. If subjects choose the algorithm, they are fully committed to the algorithm's decisions for the entire supergame. Importantly, even if participants choose to delegate pricing to the algorithm, they still proceed through each period of the supergame using the same interface as in \textsc{Baseline}. If the algorithm is adopted, its price is pre-selected and displayed on the decision screen. Participants then move through each round individually. This ensures that the time spent and interaction structure remain consistent across treatments.

In the \textsc{Recommendation} treatment, subjects can choose to get support from the algorithm, also once at the beginning of each supergame. Unlike \textsc{Outsourcing}, the algorithm's choices are only recommendations. Participants can still set prices as they wish, that is, they can overwrite the recommendation in each round of the supergame. If they decide to adopt it, the algorithm's suggested price is pre-filled in the price selection field and clearly indicated as a recommendation, allowing participants to either accept it or enter a different value.\footnote{We provide screenshots of the decision screens in online appendix Figures \ref{fig:decision_screen_out} and \ref{fig:decision_screen_rec}.}

\subsection{Algorithm}

We use Q-learning algorithms that belong to the class of reinforcement learning algorithms \citep{Watkins1989LearningRewards, Watkins1992Q-Learning}. Q-learning algorithms are designed to solve Markov decision processes with an ex-ante unknown environment. In other words, the algorithm learns everything by itself without being instructed to follow a particular strategy and without having prior knowledge of the market environment. Its objective is to maximize the stream of future discounted rewards. The algorithms learn how to price the product in a given market environment by interacting with other algorithms in a simulated market environment. It is the primary building block of the popular reinforcement learning algorithms that learn to play video games or beat humans at the game of Go \citep[e.g.,][]{mnih2015human, silver2016mastering}. Also, there is evidence that companies use those algorithms for their pricing systems \citep[see, for instance,][]{liu2019dynamic, chen2023reinforcement}.\footnote{Similarly, self-learning algorithms are increasingly popular among third-party algorithm providers, which offer pricing software to less sophisticated sellers on e-commerce platforms like Amazon. For instance, \citet{calzolari2025pricing} document that 27\% of repricer companies report that their algorithms are self-learning.}

Q-learning has been used to study the behavior of pricing algorithms \citep[see, for instance,][]{calvano2020artificial, abada2023artificial, klein2021autonomous, johnson2020platform, martin2025spoils}. These studies show that the algorithms can learn to collude on non-competitive prices tacitly. A firm that wants to outsource its pricing decision to an algorithm might be incentivized to use such algorithms, as it can facilitate market coordination and thereby increase profits. As such, it is a natural choice for our experiment. We follow the approach by \citet{werner2021algorithmic} to train the algorithm. As the market environment we consider is the same as in this paper, the algorithms also converge to the same collusive outcomes. Thus, we consider the exact same algorithm in our experiment. 

The algorithm learns a win-stay lose-shift strategy (WSLS) or perfect tit-for-tat, popularized for the iterated prisoners' dilemma by \citet{Nowak:1993uv}. It cooperates at the monopoly price if both firms picked the monopoly in the previous round. If the competitor deviates, it punishes this deviation by playing the stage game Nash equilibrium price of one. If both firms play the Nash equilibrium in a given round, the algorithm reverts to the monopoly price of $p^{M}=4$. For any other combination of prices from the previous period, it chooses the stage game Nash equilibrium price of 1 again.

While humans rarely adopt win-stay lose-shift strategies in strategically similar environments \citep[see, for instance,][]{Wright2013PunishmentMarkets, dal2019strategy}, the strategy has several advantages. By adopting it, firms gain a tool that actively promotes collusion rather than simply imitating the behavior they would otherwise follow.  First, the algorithm punishes deviations and makes collusion incentives compatible. Competitors are better off colluding with the algorithm than deviating in the current round for some immediate additional profit, but at the expense of punishment afterward. Furthermore, the algorithm is forgiving. Strategies like the grim trigger can also make collusion incentive compatible, but they can never return to the collusive price level. This outcome is undesirable for an algorithm in a pricing environment because competitors may explore the action space or make mistakes, and a trigger strategy would perceive these actions as deviations. While the algorithm has these advantages, it is important to highlight that they arise from the learning process of the algorithm and were not designed by us as the experimenter.

The algorithm is the same for all participants, and we only use its limiting strategy after convergence in a simulated training environment. That is, the algorithm does not learn during the experiment but applies the strategy it previously developed through self-play against other algorithms. This design choice reflects how reinforcement learning is often deployed in practice. Q-learning learns slowly, which makes real-time learning in actual market settings impractical. Moreover, training through self-play is a standard and effective method in reinforcement learning, as it yields robust strategies that generalize well across a variety of opponents, which is particularly important when firms do not know whether their competitors are humans or algorithms \citep[see also][]{calvano2020artificial}.

Importantly, our design is not specific to the use of self-learning algorithms. The strategy to which it converges is a one-period punishment rule that rewards mutual cooperation and punishes a deviation for a single period. This is the same kind of rule a firm might program directly to sustain collusion. In this sense, our setup captures a broader class of algorithmic behavior where firms either discover or explicitly implement strategies that make collusion incentive-compatible. What matters is not how the strategy is found, but that it reflects a credible threat to punish deviations and a willingness to return to cooperation.

While in practice, the algorithms used by competing firms do not necessarily have to be the same, it can often be the case. For instance, \citet{harrington2022effect, harrington2024hub} argues that pricing algorithms are frequently supplied by a common intermediary, meaning that competing firms often rely on the same or very similar algorithms.\footnote{For example, in Cornish-Adebiyi v. Caesars, the U.S. Department of Justice argued that competing hotels delegated their pricing to the same algorithmic provider \citep{doj2024cornish}.} As such, the setup of firms using the same algorithm reflects real-world scenarios. Moreover, the strategy learned by the pricing algorithm is a type commonly learned by reinforcement learning algorithms, as shown by \citet{kasberger2023algorithmic, schaefer2022emergence, barfuss2022intrinsic}. Hence, it is reasonable to assume that firms using reinforcement learning-based algorithms ultimately rely on algorithms that share a similar strategy.

\subsection{Procedures}

The experiment was programmed in oTree \citep{chen2016otree}, and a total of 19 sessions were conducted. Of these, thirteen were conducted at the DICELab in Düsseldorf, and six sessions at PLEx in Potsdam.\footnote{We aimed to balance session sizes across treatments. Most sessions had 20 participants, but some were split into smaller sessions of 10 due to no-shows, resulting in a mix of 20- and 10-participant sessions at both locations.} Overall, 100 participants were assigned to each treatment, totaling 300 participants in the experiment.

At the beginning of the session, participants receive neutrally framed instructions on the computer screen.\footnote{We do not label pricing choices as (un)ethical. Related work shows that machine delegation can raise unethical behavior under explicit moral framing \citep{kobis2025delegation}. How such framing would interact with adoption and pricing in our setting is an open question for future research.} The instructions and additional information are also available at any time during the experiment.\footnote{See online appendix \ref{sec:instructions} for the instructions.} After reading the instructions, participants answer a series of control questions. If a participant gives an incorrect answer three times, we provide the correct answer and explain the question. 

The sessions in all treatments had five supergames. At the beginning of each session, two participants are matched for the duration of the supergames. At the start of the next supergame, participants are randomly rematched into new markets within a matching group of ten participants. The random rematching reduces the likelihood of potential reputation effects across supergames.

Participants in the \textsc{Outsourcing} and \textsc{Recommendation} treatments receive information about the algorithm at the beginning of the session and at each supergame. They know that the algorithm is self-learned and that its objective is to maximize the participant's long-run profit. To further align expectations on the possible performance of the algorithm, participants receive a comprehensive overview of the algorithm's past performance in different market compositions from the previous experiment by \cite{werner2021algorithmic}.\footnote{The experimental design in \cite{werner2021algorithmic} differs in that the participants were assigned to a condition and did not have the choice to adopt an algorithm. We inform our participants of this fact. We also emphasize that past performance is not indicative of future profits, particularly because the use of the algorithm was predetermined, unlike having the additional choice of delegating to the algorithm initially.} The algorithm's past performance is presented as a  \(2 \times 2\) payoff matrix, showing the outcomes when no firm, one firm, or both used the algorithm.

Right after the control questions, all participants play three trial supergames with a total of 15 periods against the algorithm. This allows every participant to understand how a supergame works and, in the case of participants of treatments \textsc{Outsourcing} and \textsc{Recommendation}, also to comprehend the algorithm strategy better.\footnote{To keep the learning experience the same across treatments, also in the \textsc{Baseline} treatment, participants engage in the simulation against the algorithm. Despite this additional learning opportunity, market prices in \textsc{Baseline} are similar to previous experiments that use similar environments \citep[see, for instance,][]{horstmann2018number}. While in \textsc{Recommendation} and \textsc{Outsourcing} we explicitly tell participants that they play against the algorithm, we do not provide those details in \textsc{Baseline} to avoid any confusion.} These three simulation supergames were not payoff relevant.

At the end of each period, participants receive information about prices and profits. At the end of each supergame, we elicit participants' beliefs in \textsc{Recommendation} and \textsc{Outsourcing}.  Participants are asked to state how much confidence, in percentages, they have that their opponent used the pricing algorithm in the previous supergame (see Figure \ref{fig:elicitation_screen} in the online appendix for the elicitation screen). They can receive an additional reward of 180 experimental currency units (ECU) for guessing correctly. We incentivize belief elicitation using the binarized scoring rule \citep{hossain2013binarized} with a framing similar to \citet{danz2022belief}.

At the end of the session, participants answer an end-of-study survey regarding demographics and algorithm trust. They are then informed of the reward they earn in euros, including their profit in one selected supergame, the correctness of their belief in algorithm use, a 6 Euro show-up fee, and information about how it is computed. One Euro corresponded to 140 ECU. On average, participants earned 18.61 Euro.

\subsection{Hypotheses}\label{hypotheses}

Previous experiments in non-strategic environments have shown that algorithm trust is increased when participants have some control over the final decision \citep{Dietvorst2016overcoming}. Furthermore, in \textsc{Outsourcing}, using the algorithm is possibly associated with some cost. While the algorithm can foster collusion, participants lose the ability to implement their own strategy, which might be important to them. Conversely, using the algorithm is costless in \textsc{Recommendation}. Participants can choose a price different from the one the algorithm recommends. Furthermore, besides recommending a pricing strategy, it also allows them to anticipate how their competitor might behave if they also use the algorithm. Therefore, participants are more likely to delegate to pricing algorithms when they can override the algorithm's pricing decision than when they must completely delegate it. 

\begin{hypothesis}\label{hypo:delegation}
    Participants are more likely to delegate in \textsc{Recommendation} than in \textsc{Outsourcing}.
\end{hypothesis}

The algorithm has the potential to increase prices and profits through its deterministic collusive strategy, which incorporates a one-period punishment. Furthermore, the algorithm cannot be exploited: Any strategy other than colluding with the algorithm at $p = 4$ results in a lower profit. Participants know this both implicitly from the instructions and through practice rounds before the first supergame. If subjects delegate their decisions to the algorithm or follow its recommendations, prices will be higher. In turn, a propensity to delegate or follow recommendations will likely stem from an intention to maximize payoffs. Thus, we expect prices to be higher in \textsc{Outsourcing} and \textsc{Recommendation} than in the baseline treatment.

\begin{hypothesis}\label{hypo:prices}
    In \textsc{Outsourcing} and \textsc{Recommendation}, prices are higher compared to \textsc{Baseline}.
\end{hypothesis}

An open question is whether the ability to override the algorithms leads to higher prices (\textsc{Recommendation}) compared to fully delegating pricing decisions to an algorithm (\textsc{Outsourcing}). On the one hand, we hypothesize that participants delegate more when they can override the pricing decision. On the other hand, conditional on the decision to use an algorithm, the opportunity to override the recommendation of the collusive algorithm may lead to lower market prices compared to fully outsourcing the pricing decision. As an exploratory research question, we ask which of these two effects dominates. 

Further, comparison between and across treatments allows us to determine whether the level of commitment to the algorithm decision comes into play when deciding whether to use it. While the \textsc{Baseline} treatment includes only humans, in the case of \textsc{Outsourcing} and \textsc{Recommendation}, there may exist three types of markets: fully human markets (HH), fully algorithmic markets (AA), and mixed markets (AH). Comparison between market types helps determine whether one of them is more collusive.

\section{Results} \label{sec:results}
\subsection{Algorithm adoption}\label{sec:algorithm_adoption}

\begin{figure}[t]
        \centering
        \includegraphics[width=0.8\linewidth]{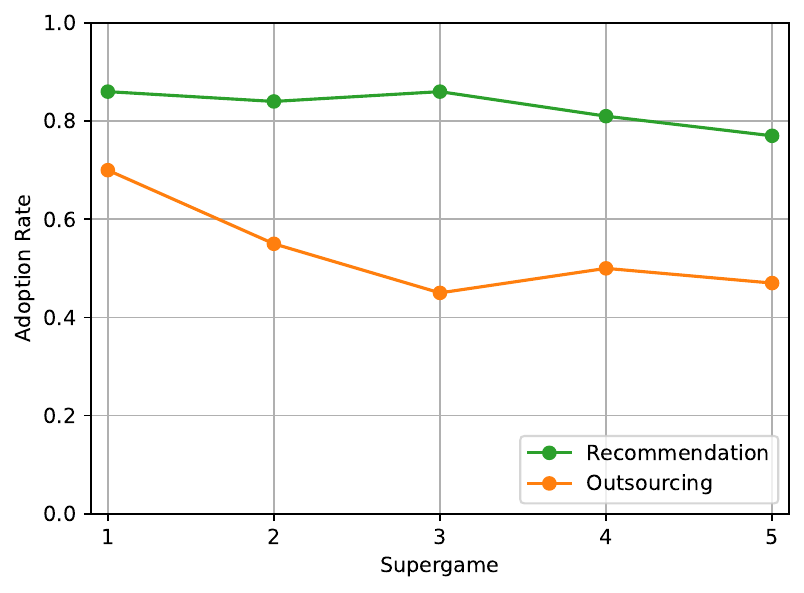}
        \caption{Algorithm adoption rates across supergames.}
        \label{fig:delegation}
\end{figure}

Figure \ref{fig:delegation} shows how often participants delegate their decisions to the algorithm. Since the decision to adopt the algorithm was made once at the beginning of each supergame in \textsc{Recommendation} and \textsc{Outsourcing}, we report the average for the five supergames separately. Algorithm adoption rates range from 45\% to 86\%, depending on the treatment and supergame, indicating substantial adoption of the algorithm. Algorithm adoption rates are consistently higher in \textsc{Recommendation} compared to \textsc{Outsourcing}. While we observe high adoption rates at the start of the experiment, these decline, particularly in \textsc{Outsourcing} across supergames.

Table \ref{table:delegation} presents the results of linear regressions that test for the statistical significance of these findings. \textsc{Recommendation} has significantly higher adoption rate than \textsc{Outsourcing} (represented by the constant).\footnote{In line with our pre-registration (https://osf.io/fh89u/), we refer to results as statistically significant when $p < 0.05$ and as weakly significant when $p < 0.1$.} The data thus reject the null hypothesis and are consistent with Hypothesis \ref{hypo:delegation}, that participants are more likely to adopt an algorithm for pricing decisions in the \textsc{Recommendation} than in the \textsc{Outsourcing}.\footnote{The main regression results are corroborated by Mann-Whitney $U$ tests in Table~\ref{tab:np_mwu_main} in the online appendix.} This pattern holds consistently across all supergames. Regression (3) shows a statistically significant decline in adoption rates across supergames. The Supergame coefficient applies to \textsc{Outsourcing} and is significant. The trend in \textsc{Recommendation} is also significant (Wald test, Supergame + \textsc{Recommendation} $\times$ Supergame, $p < 0.05$).

\begin{result}\label{result:adoption}
Participants frequently delegate to the algorithm, but adoption declines across supergames. The adoption rate is higher in \textsc{Recommendation} than in \textsc{Outsourcing}.
\end{result}

\begin{table}[t]
\centering
\caption{Algorithm adoption by treatment \label{table:delegation}}
\begin{center}
\begin{tabular}{l c c c}
\toprule
 & \multicolumn{3}{c}{Algorithm Adoption} \\
\cmidrule(lr){2-4}
 & (1) & (2) & (3) \\
\midrule
Recommendation                    & $0.29^{***}$ & $0.29^{***}$  & $0.20^{***}$  \\
                                  & $(0.05)$     & $(0.05)$      & $(0.06)$      \\
Supergame                         &              & $-0.04^{***}$ & $-0.05^{***}$ \\
                                  &              & $(0.01)$      & $(0.01)$      \\
Recommendation $\times$ Supergame &              &               & $0.03^{**}$   \\
                                  &              &               & $(0.01)$      \\
(Intercept)                       & $0.53^{***}$ & $0.64^{***}$  & $0.69^{***}$  \\
                                  & $(0.04)$     & $(0.05)$      & $(0.04)$      \\
\midrule
Sub-sample & Out \& Rec  & Out \& Rec & Out \& Rec \\
$\#$ observations & 1000 & 1000 & 1000 \\
$\# $ cluster & 20 & 20 & 20 \\
R$^2$ & 0.10 & 0.11 & 0.11 \\
\bottomrule
\multicolumn{4}{p{0.8\textwidth}}{\scriptsize{Data restricted to Outsourcing and Recommendation treatments. Results are based on linear regression with clustered standard errors at the matching group level. Significance levels: *** \( p < 0.01 \); ** \( p < 0.05 \); * \( p < 0.1 \) .}}
\end{tabular}
\end{center}
\end{table}

\begin{figure}[H]
\centering
\includegraphics[width=0.8\textwidth]{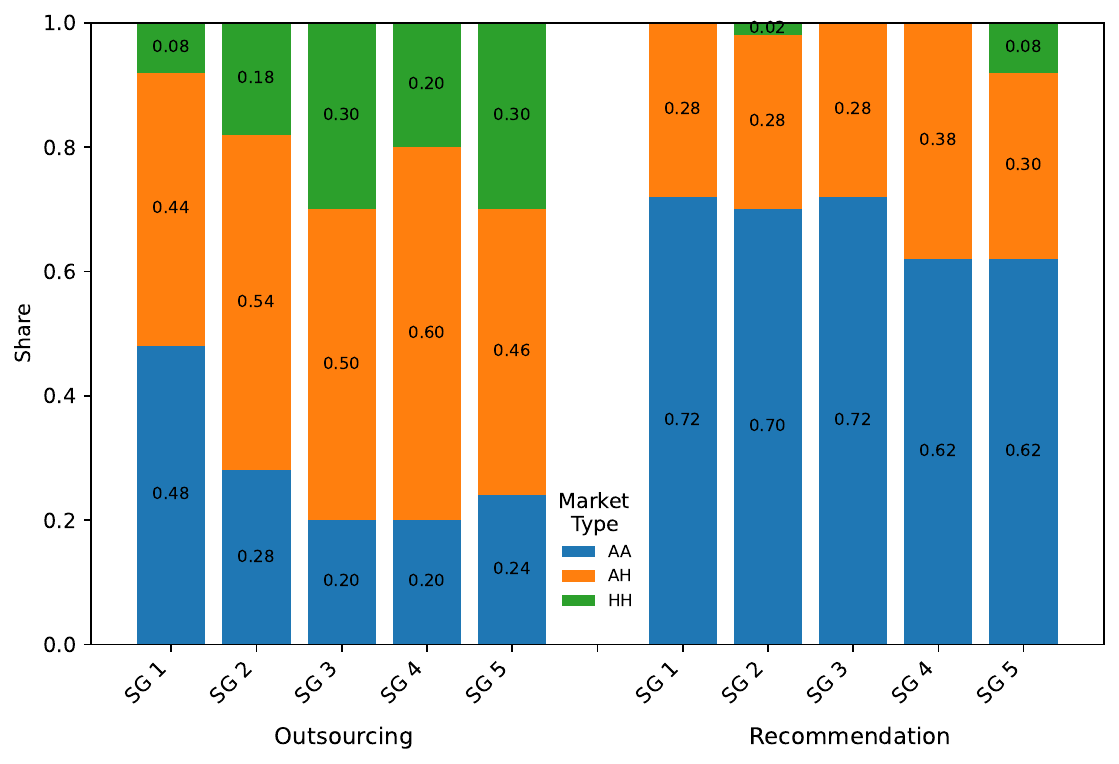}
\caption{Distribution of market types across treatments and supergames. AA refers to fully algorithmic, AH to mixed human-algorithm, and  HH to fully human markets.}
\label{fig:market_types_dis}
\end{figure}

Figure~\ref{fig:market_types_dis} illustrates how market composition varies depending on whether none, one, or both participants adopt the algorithm (``market type''). The right panel presents the data for \textsc{Recommendation} and the left panel for \textsc{Outsourcing}.  
In \textsc{Outsourcing}, mixed human-algorithmic markets dominate, accounting for around half of all observations. Fully algorithmic markets (and hence perfectly collusive ones) constitute only around one quarter of the outcomes after supergame 1. In \textsc{Recommendation}, markets in which both participants choose algorithmic support dominate with 62\% to 72\% with a moderate decline across supergames. 

Figure~\ref{fig:market_types_dis} highlights a trivial but important fact: not all markets involve algorithms acting on behalf of both players. This emphasizes the importance of human–algorithmic interaction in mixed markets. 

Concluding this section, we analyze participants' beliefs about their counterparts' algorithm use. The belief was elicited after every supergame, and we compare them to the actual delegation decisions. The accuracy of beliefs differs noticeably across treatments. In \textsc{Recommendation}, the mean belief conditional on the opponent choosing the algorithm is $59$\%, while the belief when the opponent did not is $46.4$\%. In contrast, the \textsc{Outsourcing} treatment yields a pronounced gap, with beliefs at $80.6$\% when the opponent used the algorithm and $38$\% when the opponent did not. Table  \ref{table:belief_result2} reports a metric for the accuracy of beliefs as well as tests showing that stated beliefs are better than random guesses. 

\begin{result}
Participants' beliefs about their opponent's algorithm use are significantly more accurate than chance in both treatments. Beliefs in the \textsc{Outsourcing} treatment are significantly more accurate than those in the \textsc{Recommendation} treatment. 
\end{result}

\subsection{Prices}

\subsubsection{Market prices across supergames and treatment comparison}

\begin{figure}[ht]
        \centering
        \includegraphics[width=0.8\linewidth]{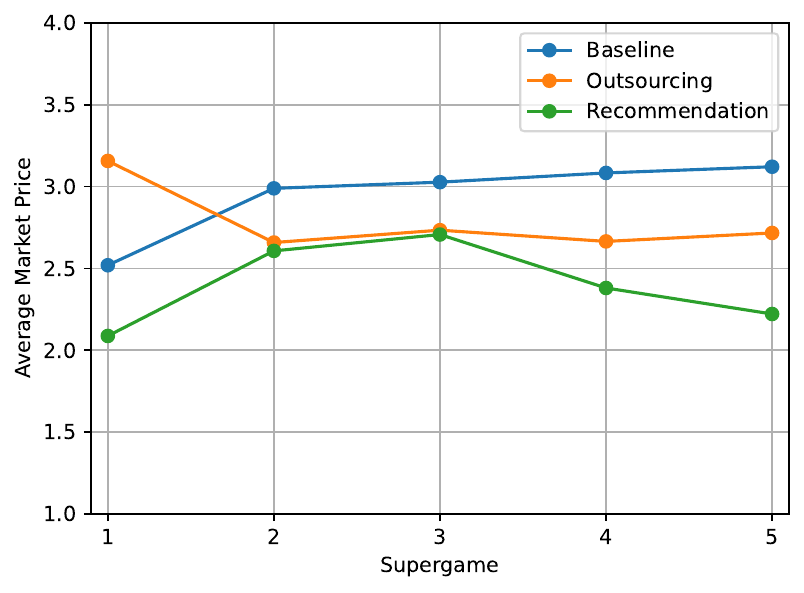}
        \caption{Average market price across supergames}
        \label{fig:wprice_main}
\end{figure}

Figure \ref{fig:wprice_main} presents the average market price across all rounds, separately for each supergame and treatment, where the market price is defined as the lower of the two prices. Average market prices range from 2.2 to 3.2, depending on the treatment and supergame. Prices in \textsc{Baseline} are higher than in the other two treatments, except in the first supergame, and market prices in \textsc{Outsourcing} are higher than in \textsc{Recommendation}.

The regressions in the top panel of Table \ref{table:market_prices_overall} analyze the market prices for treatment differences. Regression (1) confirms that the average market price in \textsc{Baseline} is significantly higher than in \textsc{Recommendation} ($p < 0.01$) and not significantly different from \textsc{Outsourcing}. These results remain robust when we weight the data by supergame length in regression (2). 
In supergame 1, market prices in \textsc{Outsourcing} are significantly higher than in \textsc{Baseline} while prices in \textsc{Recommendation} are significantly lower, see regression (3). 

By supergame 5, this pattern is reversed. As shown in regression (4), there is a significant difference in market prices between \textsc{Baseline} and \textsc{Outsourcing} in supergame 5 ($p < 0.05$).
Extending regressions (1) and (2), regressions (5) and (6) add a supergame time trend and treatment$\times$supergame interactions while keeping the same sample and weighting schemes, respectively. The positive and significant supergame coefficient confirms the upward trend in \textsc{Baseline}, whereas the negative and significant interaction terms for \textsc{Outsourcing} and \textsc{Recommendation} indicate no or declining price trends in the algorithmic treatments.
The difference in market prices between \textsc{Outsourcing} and \textsc{Recommendation} is highly significant in the full sample (regression (1), $p < 0.01$, Wald test), weakly significant in the weighted model (regression (2), $p < 0.1$, Wald test), highly significant in supergame 1 (regression (3), $p < 0.01$, Wald test), significant in supergame 5 (regression (4), $p < 0.05$, Wald test), and again highly significant in the specifications with treatment $\times$ supergame interactions (regressions (5) and (6), $p < 0.01$, Wald tests).

Market prices increase in the \textsc{Baseline} treatment across supergames, decline in \textsc{Outsourcing}, and follow no statistically significant trend in \textsc{Recommendation}. For \textsc{Baseline}, we find a highly significant and positive \textit{supergame} coefficient ($p < 0.01$) in regressions (5) and (6), suggesting a trend toward greater cooperation as participants learn over supergames.\footnote{This pattern is similar to results from other experiments with strategically similar environments with two players \citep[see, for instance,][]{dal-bo-frechette-JEL-2018}.}

\begin{table}[H]
\caption{Market prices and first-round market prices by treatment}
\label{table:market_prices_overall}
\begin{center}
\begin{tabular}{l c c c c c c}
\toprule
& \multicolumn{6}{c}{Market Prices} \\
\cmidrule(lr){2-7}
& (1) & (2) & (3) & (4) & (5) & (6) \\
\midrule
Outsourcing                                                      & $0.18$        & $-0.16$       & $0.64^{***}$ & $-0.40^{**}$  & $0.82^{***}$  & $0.49^{**}$   \\
                                                                 & $(0.15)$      & $(0.15)$      & $(0.20)$     & $(0.18)$      & $(0.23)$      & $(0.21)$      \\
Recommendation                                                   & $-0.56^{***}$ & $-0.55^{***}$ & $-0.43^{**}$ & $-0.90^{***}$ & $-0.30$       & $-0.17$       \\
                                                                 & $(0.18)$      & $(0.18)$      & $(0.21)$     & $(0.22)$      & $(0.23)$      & $(0.21)$      \\
Supergame                                                        &               &               &              &               & $0.16^{***}$  & $0.13^{***}$  \\
                                                                 &               &               &              &               & $(0.04)$      & $(0.04)$      \\
\begin{tabular}{@{}l@{}}Outsourcing  $\times$ SG\end{tabular}    &               &               &              &               & $-0.28^{***}$ & $-0.22^{***}$ \\
                                                                 &               &               &              &               & $(0.06)$      & $(0.05)$      \\
\begin{tabular}{@{}l@{}}Recommendation  $\times$ SG\end{tabular} &               &               &              &               & $-0.11^{*}$   & $-0.13^{**}$  \\
                                                                 &               &               &              &               & $(0.06)$      & $(0.05)$      \\
(Intercept)                                                      & $2.77^{***}$  & $2.95^{***}$  & $2.52^{***}$ & $3.12^{***}$  & $2.40^{***}$  & $2.56^{***}$  \\
                                                                 & $(0.08)$      & $(0.08)$      & $(0.11)$     & $(0.11)$      & $(0.14)$      & $(0.14)$      \\
\midrule
Supergame & All & All & SG 1 & SG 5 & All & All \\
$\# $ observations & 15300 & 15300 & 8400 & 3150 & 15300 & 15300 \\
$\# $ cluster & 30 & 30 & 30 & 30 & 30 & 30 \\
R$^2$ & 0.05 & 0.03 & 0.10 & 0.07 & 0.07 & 0.04 \\
\midrule
& \multicolumn{6}{c}{First Round Market Prices} \\
\cmidrule(lr){2-7}
& (1) & (2) & (3) & (4) & (5) & (6) \\
\midrule
Outsourcing                                                      & $0.33^{*}$   & $0.21$       & $0.92^{***}$ & $0.12$       & $0.93^{***}$  & $0.69^{**}$   \\
                                                                 & $(0.17)$     & $(0.17)$     & $(0.28)$     & $(0.17)$     & $(0.28)$      & $(0.29)$      \\
Recommendation                                                   & $0.39^{**}$  & $0.32^{*}$   & $0.92^{***}$ & $0.16$       & $0.97^{***}$  & $0.83^{***}$  \\
                                                                 & $(0.18)$     & $(0.18)$     & $(0.25)$     & $(0.19)$     & $(0.25)$      & $(0.26)$      \\
Supergame                                                        &              &              &              &              & $0.18^{***}$  & $0.16^{***}$  \\
                                                                 &              &              &              &              & $(0.05)$      & $(0.04)$      \\
\begin{tabular}{@{}l@{}}Outsourcing  $\times$ SG\end{tabular}    &              &              &              &              & $-0.20^{***}$ & $-0.16^{***}$ \\
                                                                 &              &              &              &              & $(0.06)$      & $(0.06)$      \\
\begin{tabular}{@{}l@{}}Recommendation  $\times$ SG\end{tabular} &              &              &              &              & $-0.19^{***}$ & $-0.17^{***}$ \\
                                                                 &              &              &              &              & $(0.06)$      & $(0.06)$      \\
(Intercept)                                                      & $2.85^{***}$ & $2.94^{***}$ & $2.38^{***}$ & $3.08^{***}$ & $2.31^{***}$  & $2.43^{***}$  \\
                                                                 & $(0.14)$     & $(0.14)$     & $(0.21)$     & $(0.15)$     & $(0.22)$      & $(0.21)$      \\
\midrule
Supergame & All & All & SG 1 & SG 5 & All & All \\
$\# $ observations & 750 & 750 & 150 & 150 & 750 & 750 \\
$\# $ cluster & 30 & 30 & 30 & 30 & 30 & 30 \\
R$^2$ & 0.03 & 0.02 & 0.16 & 0.01 & 0.05 & 0.03 \\
\bottomrule
\multicolumn{7}{p{0.95\textwidth}}{\scriptsize{Results are based on linear regression with standard errors clustered at the matching group level. Models (2) and (6) are weighted by supergame length. Significance levels: *** \( p < 0.01 \); ** \( p < 0.05 \); * \( p < 0.1 \).}}
\end{tabular}
\end{center}
\end{table}

In contrast, average market prices in the algorithmic treatments fluctuate across supergames, with an overall negative trend across supergames in \textsc{Outsourcing} (regressions (5) and (6), Supergame + Supergame $\times$ \textsc{Outsourcing}, $p < 0.01$, Wald tests), and no statistically significant trend in \textsc{Recommendation} (regressions (5) and (6), Supergame + Supergame $\times$ \textsc{Recommendation}, $p > 0.1$, Wald tests).

\begin{result}\label{result:market_prices}
Market prices are higher in \textsc{Baseline} than in \textsc{Recommendation} and \textsc{Outsourcing} in the final supergame.
\end{result}

Result \ref{result:market_prices} stands in sharp contrast to Hypothesis \ref{hypo:prices}. Even though the algorithm available to participants implements highly collusive strategies and makes collusion incentive compatible, it does not foster collusion. Only in the first supergame \textsc{Outsourcing} has a pro-collusive effect. In the last supergame, access to the algorithm even has a pro-competitive effect and leads to lower prices than when no algorithm can be used at all.

\subsubsection{Market prices in the initial periods} \label{sec:market_price_periods}

\begin{figure}[hbt!]  
\centering         
\includegraphics[width=0.9\linewidth]{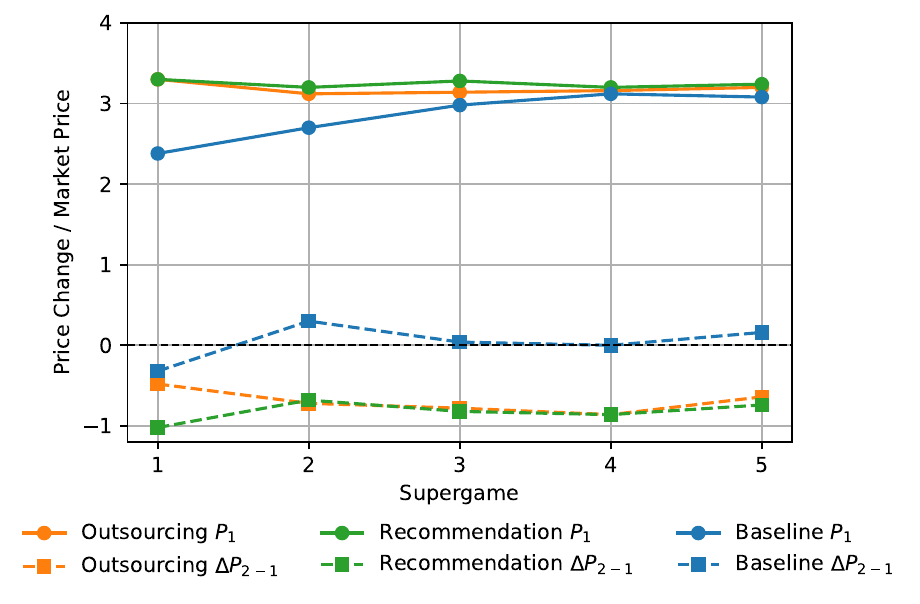}
\caption{Average first round market price $P_1$ and difference between second and first round $\Delta P_{2-1}$}\label{fig:price_diff_round1v2} 
\end{figure}

Next, we examine the market prices in the first periods of the supergames, which are central to many collusive strategies and often predictive of cooperative outcomes \citep[see, for instance,][]{dal-bo-frechette-JEL-2018}. The algorithm either sets or recommends a price of 4 in the first round for adopters, while participants in the \textsc{Baseline} treatment and non-adopters freely choose their initial market price. The three solid lines in Figure \ref{fig:price_diff_round1v2} visualize the average first-round market prices across supergames. 

The bottom panel of Table \ref{table:market_prices_overall} analyzes the first-round market prices for treatment differences. Regression (3) shows that first-round market prices in both \textsc{Outsourcing} and \textsc{Recommendation} are higher than in \textsc{Baseline} ($p < 0.01$). By supergame 5, these differences are no longer significant (regression (4), $p > 0.1$). In the full sample, first-round prices in the algorithmic treatments exceed \textsc{Baseline} (regression (1)), and the effects weaken under weighting (regression (2)). Adding a time trend and interactions in regressions (5)–(6) yields a positive Supergame coefficient and negative treatment $\times$ supergame terms, indicating convergence toward similar initial prices. Wald tests do not reject equality between \textsc{Outsourcing} and \textsc{Recommendation} in any specification (regressions (1)–(6), $p > 0.1$), nor do they reject equality of the interaction effects (regressions (5) and (6), $p > 0.1$). Within-treatment trends are positive in \textsc{Baseline} but not statistically different from zero in the algorithmic treatments (regressions (5) and (6), $p > 0.1$).

The three dashed lines in Figure \ref{fig:price_diff_round1v2} show the market price change from period 1 to period 2. This change is negligible in \textsc{Baseline} across all supergames, whereas prices drop by a substantial amount (on average 0.5 to 1) in the algorithmic treatments. That is, participants in \textsc{Baseline} maintain the initial price level, unlike in \textsc{Outsourcing} and \textsc{Recommendation}, where prices quickly decline.\footnote{This sharp decline in prices is not limited to the first two rounds but reflects a broader pattern. Average Market prices are generally below first-round market prices in the algorithmic treatments. By contrast, in the \textsc{Baseline} treatment, we do not find a significant difference as shown in Table~\ref{tab:C04_first_round_effect_by_treatment} in the online appendix.}

\begin{figure}[hbt!]
        \centering
        \includegraphics[width=0.95\linewidth]{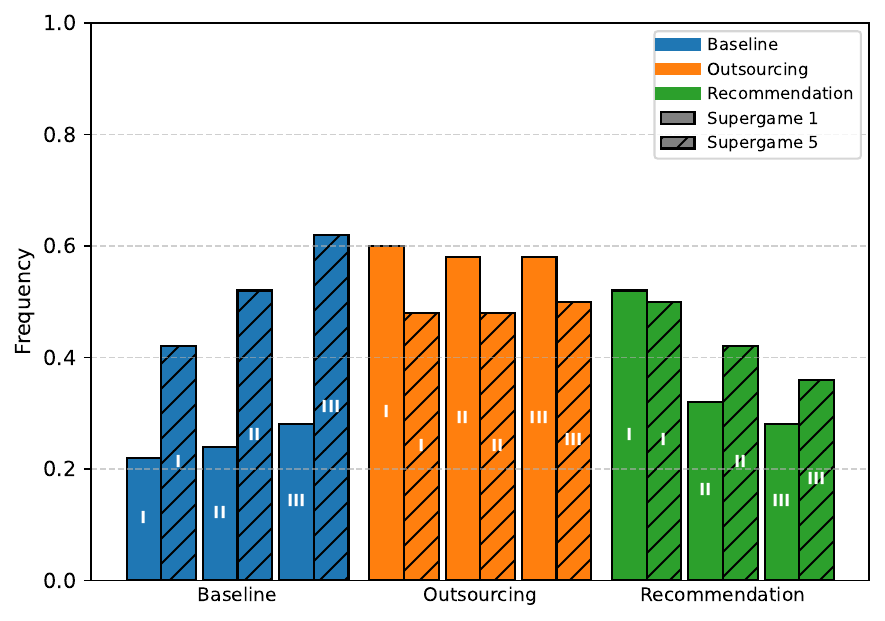}
        \caption{Frequency of joint monopoly pricing outcomes ($p_1 = p_2 = 4$) in the first three rounds (I, II, III) of Supergames 1 (solid bars) and 5 (hatched bars), by treatment.}
\label{fig:outcome_distribution_first3}
\end{figure}

The pattern of failed coordination in the algorithmic treatments versus improved coordination in \textsc{Baseline} is also evident in Figure \ref{fig:outcome_distribution_first3}. It shows the frequency of outcomes, where both firms choose the monopoly price, across the first three rounds of supergames 1 and 5. In \textsc{Baseline}, the frequency of these collusive outcomes increases steadily, especially by supergame 5, indicating successful learning and coordination. In contrast, the share of fully collusive outcomes in \textsc{Outsourcing} remains flat, and in \textsc{Recommendation} it even declines across the first three rounds. This suggests that while humans in \textsc{Baseline} learn to coordinate on fully collusive outcomes over time, algorithms fail to improve this pattern, and in the case of \textsc{Recommendation}, may actively hinder it.\footnote{See Figure \ref{fig:outcome_distribution_first3_all} in the online appendix for the frequency broken down by other price states.}

\begin{result}\label{result:decline}
While first-round market prices in \textsc{Outsourcing} and \textsc{Recommendation} are higher than in \textsc{Baseline}, they increase in \textsc{Baseline} and are not statistically significantly different across treatments by the last supergame. 
\end{result}

Result \ref{result:decline} suggests that while algorithms help establish initially high price levels, they also introduce immediate downward pressure, often leading to declining prices across rounds, in particular in \textsc{Recommendation}. This contrasts with the \textsc{Baseline} treatment, where we observe rising prices over time. The pattern is intuitive. Perfect collusion is rare in early rounds, triggering algorithmic punishment, which participants tend to follow.  However, participants do not seem to sufficiently learn to return to the collusive path after punishment, which is recommended by the algorithm.

\section{Discussion} \label{sec:discussion}

While subjects choose the algorithm, they do so only partially and at a decreasing rate. Furthermore, contrary to Hypothesis \ref{hypo:prices}, after learning, market prices are higher in the \textsc{Baseline} treatment, where two humans interact without the assistance of an algorithm. While the initial prices in \textsc{Recommendation} and \textsc{Outsourcing} are indeed collusive, they decrease in the following periods. In the following, we discuss some of the mechanisms that drive those results.

\subsection{Coordination failure at the stage of algorithm choice}

If both participants choose the algorithm or follow its recommendations, they automatically monopolize the market. However, Section \ref{sec:algorithm_adoption} documents that participants do not always coordinate on using an algorithm. In this section, we discuss the consequences of this coordination failure.

The different market compositions seen in Figure \ref{fig:market_types_dis} lead to distinct price outcomes across treatments seen in Figure \ref{fig:wprice_mkttype} in the online appendix. In \textsc{Outsourcing}, AA markets consistently achieve the monopoly price of 4, as the algorithm starts by colluding and never deviates. Mixed markets show lower prices, despite the fact that participants could maintain collusion by also choosing the monopoly price in each round. Prices in HH markets are the lowest, even lower than in \textsc{Baseline}, although both involve two humans freely choosing prices. 
In \textsc{Recommendation}, prices in AA markets are lower than in \textsc{Outsourcing}, averaging around 2.5. This suggests that participants do not consistently follow the algorithm's recommendations. Market prices in AH are slightly lower still, indicating limited coordination. Overall, market composition strongly shapes pricing outcomes, with algorithmic coordination failing to match the stability and collusion observed in \textsc{Outsourcing}.

\begin{table}[hbt!]
\caption{Per-period profits, average across five supergames, \textsc{Outsourcing} and \textsc{Recommendation} treatments}
\label{tab:normalized_profits}
\centering

\begin{tabular}{lcc}
\multicolumn{3}{c}{\sc Outsourcing} \\
\toprule
 & Algorithm & No Algorithm \\
\midrule
Algorithm & 120.00, 120.00 & 57.46, 95.22 \\
No Algorithm & ~~95.22, ~57.46 & 50.34, 50.34 \\
\bottomrule
\end{tabular}

\bigskip

\begin{tabular}{lcc}
\multicolumn{3}{c}{\sc Recommendation} \\
\toprule
 & Algorithm & No Algorithm \\
\midrule
Algorithm & 68.66, 68.66 & 60.77, 61.49 \\
No Algorithm & ~~61.49, ~60.77 & 70.00, 70.00 \\
\bottomrule
\end{tabular}

\end{table}

Table \ref{tab:normalized_profits} summarizes how algorithm adoption and subsequent pricing decisions affect profits. It reports the (ex-post realized) average profits per round across the five supergames, conditional on the own adoption decision and the adoption decision of the market opponent. For \textsc{Outsourcing} (top panel of Table \ref{tab:normalized_profits}), choosing the algorithm is the dominant action, and mutual adoption would be the Nash equilibrium of this reduced game. In \textsc{Recommendation} (bottom panel), both mutual adoption and mutual non-adoption would be Nash equilibria of the reduced game, although the latter is based on relatively few observations.\footnote{We provide the same analysis split by the first and last supergame in Table \ref{tab:normalized_profits_sg1_sg5} in the online appendix. } Hence, adoption would be a Nash equilibrium outcome in both treatments and should be in the participants' best interest. 

It is important to recognize, however, that the averages reported in Table~\ref{tab:normalized_profits} mask individual-level heterogeneity. In particular, some participants may have been better off deviating from algorithm use, especially when they were matched with a human in mixed markets.\footnote{In their analysis of the price algorithms in German gas stations, \citet{assad2020algorithmic} find significantly higher margins when both firms in a duopoly use an algorithm. When only one station employs an algorithm, margins are similar to those under human pricing.} For example, in \textsc{Outsourcing}, 59.1\% of algorithm adopters paired with a human in the first supergame earned profits below the average human-human payoff. In later supergames, this share remains high, ranging from 43.5\% to 76.0\%. Even when not strictly worse off, adopters may have been discouraged by seeing their opponent earn significantly more, up to two-thirds more, which could reduce their willingness to adopt again. These patterns indicate that mixed markets involve considerable heterogeneity in individual payoffs, which can shape adoption incentives in ways not captured by aggregate averages. In the following, we investigate what drives this individual-level heterogeneity in mixed markets, starting with a closer look at the \textsc{Outsourcing} treatment.

\subsection{Failure to collude in Outsourcing}

The large payoff differences between adopters and non-adopters in asymmetric markets in \textsc{Outsourcing} (see Table~\ref{tab:normalized_profits}) can be traced to a cyclic-deviation strategy used by many non-adopters. Recall that the participants had the opportunity to explore the algorithm in advance, so they understand its behavior and can recognize when it is being used. Some participants start by checking whether their opponent has chosen the algorithm, and if so, use a strategy that involves undercutting the algorithm's collusive price once, then immediately switching to the punishment price of 1 to reset cooperation. By repeating this pattern of triggering punishment by deviating, and then playing $p=1$, participants attempt to extract short-term gains while staying synchronized with the algorithm's punishment strategy. As shown in Figure \ref{fig:avgprices_round_sg1vs5} in the online appendix, this leads to a distinct zig-zag pattern in supergame 5 in the \textsc{Outsourcing} treatment.\footnote{The zig-zag pattern is still visible, even though Figure \ref{fig:avgprices_round_sg1vs5} is based on average prices from all interactions, including duopolies that do not employ this strategy, or that employ it one period later. } 

In mixed AH markets, human participants behave as follows. Only $26.6$\% of non-adopters consistently cooperate with the algorithm, whereas the cyclic-deviation strategy is used in 34.6\% (44) out of $127$ mixed markets. When followed consistently, this strategy yields a theoretical average payoff of 15 per round for the algorithm adopter, alternating between 0 in the undercutting round and 30 during the punishment phase. Empirically, algorithm adopters facing such cycles earned just 21.1 per round on average, compared to 90.4 for those in markets where the strategy was not used. 

The prevalence of this cyclic-deviation strategy, in turn, deters subjects from adopting the algorithm in subsequent supergames. While both players are worse off than if both players adopted, the adopter of the algorithm suffers greater losses compared to the undercutter. In such cases, a participant who experiences being undercut can earn significantly less than they would in a human-human market, making non-adoption appear rational from an ex-post perspective. 

Of course, using the seemingly exploitative strategy is not profitable. Collusion at the monopoly price is incentive compatible, and the punishment strategy of the algorithm makes it impossible to achieve higher profits from deviating than from consistently colluding.\footnote{\cite{duffy2022facing} observe similar patterns in the context of a repeated Prisoner's Dilemma even when participants fully understand the algorithm and its strategy (a known grim trigger opponent). Their design eliminates strategic uncertainty and other-regarding preferences, yet most subjects make systematically suboptimal choices and deviate too frequently, similar to our results in mixed markets.} For $\delta=0.95$, the deviation payoff ($3\cdot 60$) plus punishment payoff ($\delta \cdot 1 \cdot 30$) is strictly smaller than the collusive payoff ($30 \cdot 4 \cdot (1+\delta$)). 

Participants know that colluding with the algorithm is in their best interest. They receive detailed information in the instructions about the algorithm's purpose to maximize their profits and its performance in previous experiments. In addition, they are trained against the algorithm's strategy in trial rounds. Despite this, many participants are willing to forego profits to repeatedly undercut the algorithm and make the other participant who chooses to use it worse off. In other words, we argue that this behavior is not driven by myopic decision-making but by a preference for outperforming the opponent.\footnote{The pattern that humans express spiteful preferences towards algorithms or humans that use algorithms in strategic games has also been reported for other environments \citep[see, for instance,][]{karpus2021algorithm, march2019behavioral}.}

While the price decline in mixed markets in \textsc{Outsourcing} can be attributed to the cyclic-deviation strategy, this mechanism alone cannot explain why profits are also substantially lower in the fully human (HH) markets of \textsc{Outsourcing}. These lower profits are presented in Table~\ref{tab:normalized_profits}, and they persist across supergames. There are three complementary explanations for the low prices and payoffs in HH markets. First, there are selection effects. Even in the first supergame, prices in HH markets are significantly lower than the levels observed in \textsc{Baseline}, despite being structurally identical in setup ($p<0.05$, OLS).\footnote{See Figure~\ref{fig:wprice_mkttype} for a comparison of the different market types to the \textsc{Baseline}. Table~\ref{tab:market_prices_sg1_baseline_outHH} in the online appendix provides a comparison of market prices in supergame 1 of the \textsc{Baseline} and \textsc{Outsourcing} treatments, with a focus on human-human markets.} Second, and as previously mentioned, non-adopters in \textsc{Outsourcing} often begin a supergame by checking whether the rival has chosen the algorithm. This may spoil the scope for further collusion during the remainder of the supergame. Third, prior exposure to the cyclic-deviation strategy, whether as the initiator or the counterpart, can reduce willingness to cooperate and result in more competitive pricing.

\subsection{Failure to follow recommendations of the algorithm}

In \textsc{Recommendation}, a substantial share of participants adopt the algorithm. However, prices remain consistently lower than in \textsc{Baseline}. This is striking: If recommendations were followed, markets would collude, and if ignored, prices should resemble those in \textsc{Baseline}. To investigate the mechanism behind this pro-competitive effect, we analyze the length of the punishment recommendations and the extent to which participants follow or deviate from them. 

The algorithm recommends punishment until both players jointly play the price of $p=1$. If participants generally follow recommendations with only occasional mistakes, punishment phases would last approximately one round. However, we observe that they often persist substantially longer.\footnote{We define a punishment phase as the number of consecutive rounds in which the algorithm recommends the price of $1$. The phase ends when players set a price of $p=1$ and the algorithm subsequently recommends or plays $4$. It is not required to actually switch back to a price of 4 for the recommended punishment phase to conclude.} Across all market compositions, the average recommended punishment length in \textsc{Recommendation} is $1.91$. 
This pattern is especially pronounced in AH markets, where the average punishment phase lasts 3.06 rounds (See Table~\ref{tab:punishment_durations_AA_AH} in the online appendix). The first punishment phase recommendation is particularly long, with a mean of 2.01 rounds and a heavily right-skewed distribution, with markets that remain with a punishment recommendation for up to 21 rounds (See Table~\ref{tab:first_punishment_phase} in the online appendix).\footnote{We also find that most participants in the \textsc{Recommendation} treatment do not follow the first recommendation after a punishment phase ends as shown in online appendix Table~\ref{tab:compliance_after_punishment}}

These prolonged punishment phases are driven by frequent deviations from the algorithm's price recommendations in \textsc{Recommendation}, with downward deviations occurring more often than upward ones. To illustrate this asymmetry, we divide the data by the only two recommendations the algorithm gives: either $p = 1$ or $p = 4$. When $p = 1$ is recommended, participants frequently deviate upward, with rates between 32\% and 43\% across supergames. In contrast, when $p = 4$ is recommended, adherence is higher, ranging from 50\% to 79\% (See Figure~\ref{fig:deviation_shares_recommendation_1_4} in the online appendix). These patterns help explain the prolonged punishment phases. Repeated upward deviations from $p = 1$ delay the algorithm's return to recommending $p = 4$. The same holds in terms of magnitude: downward deviations from $p = 4$  average $-2.21$, while upward deviations from $p = 1$ average $1.66$, for an absolute difference of $0.56$ (see Table~\ref{tab:mean_deviations}). Overall, the recommendations appear to effectively push prices down rather than resume the collusion after a deviation.

\begin{table}[H]
\caption{Mean deviations from algorithmic recommendation per supergame and overall}
\label{tab:mean_deviations}
\centering
\begin{tabular}{lccc}
\toprule
\shortstack{{Supergame}\\\textbf{ }} & \shortstack{{Recommendation} \\ {of 1}} & \shortstack{{Recommendation} \\ {of 4}} & \shortstack{{Absolute} \\ {Difference}} \\
\midrule
1 & 1.52 & -2.55 & 1.03 \\
2 & 1.88 & -1.92 & 0.04 \\
3 & 1.73 & -1.77 & 0.04 \\
4 & 1.67 & -2.23 & 0.56 \\
5 & 1.49 & -2.61 & 1.12 \\
\midrule
{Overall} & {1.66} & {-2.21} & {0.56} \\
\bottomrule
\end{tabular}
\end{table}

It is helpful to make a comparison to \textsc{Baseline}. In \textsc{Baseline}, participants gradually learn to collude, causing prices to rise across the initial rounds of a supergame and, ultimately, across supergames (see Section \ref{sec:market_price_periods} and Figure \ref{fig:outcome_distribution_first3}). The algorithmic recommendation seems to counteract the learning effect in \textsc{Baseline}. As a result, market prices in \textsc{Recommendation} are lower than in \textsc{Baseline}, despite the fact that the algorithm is collusive and adoption rates are high.

The tendency to override recommendations, especially those involving punishment, may reflect a form of algorithm aversion. \citet{prahl2017understanding} show in a forecasting experiment that a single visible error significantly reduces reliance on algorithmic advice, while a similar human error has little effect. In our setting, when a recommendation is perceived as unprofitable (for example, when a competitor undercuts and wins the market), it appears to trigger a similar loss of trust. Although algorithm adoption remains high, participants often discount its specific recommendations, override them, and prolong punishment phases. Evidence from \citet{Dietvorst2016overcoming} supports this interpretation. They find that allowing users to adjust an algorithm's output leads to frequent changes that lower forecast accuracy, even as usage increases. This suggests users may trust the algorithm overall but still reject individual suggestions. This pattern mirrors the behavior observed in \textsc{Recommendation}, where participants adopt the algorithm but frequently override its advice. These selective overrides prolong punishment phases and intensify downward price pressure. The result is a form of advice-specific algorithm aversion in which participants accept the algorithm in principle but reject its unadjusted recommendations in practice, especially when following them appears costly.

\section{Conclusion} \label{sec:conclusion}
Although recent studies have shown that reinforcement learning algorithms can learn to collude in repeated pricing games \citep{calvano2021algorithmic,klein2021autonomous}, little attention has been given to the fact that using such algorithms is a strategic decision made by humans. This omission is significant because humans may be reluctant to adopt such algorithms, even when they perform well. The well-documented aversion to the use of algorithms in other contexts \citep{dietvorst_algorithm_2015} could offset or even reverse the anti-competitive effects of algorithmic pricing, which is a growing concern among regulators and researchers. 

We address this gap by analyzing participants' decisions to adopt a self-learned pricing algorithm that is highly collusive when matched against itself in a laboratory experiment. Participants interact in a repeated Bertrand pricing game. We vary whether participants can adopt the algorithm, and whether adoption entails full delegation or manual override. Our design enables us to examine algorithm adoption and its effect on market prices. 

Our results show that the adoption of algorithms is substantial and consistently higher when participants can override them. However, adoption rates decline over time in both algorithmic treatments. We find no evidence that access to a collusive algorithm leads to higher market prices. In fact, prices in both treatments fall across supergames, and by the final supergame, they are significantly lower than in the human-only baseline. The mere presence of a collusive algorithm can result in more competitive outcomes when the use of an algorithm is endogenous. 

We identify distinct behavioral mechanisms for our two treatments. In the condition where subjects are fully committed to the algorithm's decisions (\textsc{Outsourcing}), not all participants adopt the algorithm. Those who retain manual control often respond with aggressive pricing strategies that reduce the earnings of algorithm users, even though exploiting the algorithm is not profitable. This reduces algorithm users’ profits, discourages further adoption, and pushes market prices down over time. Due to the ample learning opportunities and feedback provided throughout the experiment, these effects are unlikely to result from a misunderstanding of the algorithm. Nor can they be easily avoided by alternative algorithms with similar collusive capabilities, as frequent deviations are also observed in settings where algorithms use harsher punishment strategies, such as grim trigger strategies \citep{duffy2022facing}. Instead, our results suggest that the observed pattern reflects a preference for outperforming others rather than short-sighted decision-making, as documented in other experiments \citep[][]{karpus2021algorithm,march2019behavioral}, where participants are also willing to incur losses to exploit an algorithm.

In the treatment where the participants can override the algorithm's decisions (\textsc{Rec\-om\-men\-da\-tion}), adoption rates are high. However, participants frequently override the algorithm’s collusive suggestions. These deviations trigger punishment phases that often persist due to insufficient coordination or reluctance to return to the recommended path. This creates sustained downward pressure on prices. Here, too, an algorithm that can sustain collusion in self-play ends up reinforcing competition when interacting with human users. 

Overall, our results are surprising. Since the algorithm fosters collusion and its use is optional, one might expect its availability to raise prices or have little effect. For example, participants may choose not to adopt it or ignore its advice. Contrary to this expectation, the reverse occurs. Systems that reliably collude with each other can behave very differently when humans make adoption and execution decisions. In our setting, this not only weakens collusion but also reverses it. Ultimately, algorithm availability leads to lower prices. This shows the importance of studying hybrid human–algorithm environments rather than simulations involving only fully autonomous agents \citep{tsvetkova2024human}. 

From a competition policy perspective, our findings suggest caution in assuming that pricing algorithms necessarily facilitate collusion. In our experiment, we give collusion every opportunity to succeed by using a highly collusive, non-exploitable algorithm that has already learned how to sustain supra-competitive prices. Algorithms that would need to learn from scratch during the experiment are likely to perform worse, not better. The absence of collusion in our setting is not due to insufficient strategic sophistication or slow convergence.\footnote{A common critique of reinforcement learning in pricing is that it requires too much time to learn collusive strategies, making this form of algorithmic collusion unlikely in practice \citep[see, for example,][]{den2022artificial}.} Instead, it stems from behavioral factors on the human side. These dynamics, such as algorithm aversion, strategic undercutting, or spite, deserve more attention in future research to better inform both competition policy and regulation when discussing algorithmic pricing.

\pagebreak

\FloatBarrier

\bibliographystyle{aer}

\bibliography{refs.bib}
\pagebreak

\newcommand{\onlineappendixsection}[1]{%
    \refstepcounter{section} 
    \section*{Online Appendix \Alph{section}: #1} 
    \addcontentsline{toc}{section}{Online Appendix \Alph{section}: #1} 
}
\appendix
\counterwithin{figure}{section}
\counterwithin{table}{section}
{
\setlength{\parindent}{0pt}
\onlineappendixsection{Instructions} \label{sec:instructions}
\subsection*{Instructions - English translation}

In this experiment, you will make repeated decisions. These decisions allow you to earn real money. How much you earn depends on your decisions and those of the other participants. \textbf{Independently of this, you will receive \texttt{6} euros for participating.} \

In the experiment, we use a fictitious currency called talers. After the experiment, the talers will be converted into euros and paid out to you. \textbf{Here, \texttt{140} talers correspond to one euro.} \

In this experiment, you represent a firm in a virtual product market. In each market, one other firm sells the same product as you. This firm is represented by another participant. All firms offer \texttt{60} units of the comparable product. Producing the product entails no costs for the firms. The experiment consists of several periods. A period comprises several rounds, with the exact number determined randomly. You play the game for five periods. In each round of a period, you face the same firm (i.e., one participant). After each period, you are randomly paired with a new firm.

\paragraph{Round:} A round is a part within a period in which you interact with the same participant. In every round you make price decisions and receive information about your profit and the prices of the other firms.

\paragraph{Period:} A period consists of several rounds and represents a complete repetition of the game. Before a new period begins, the market is randomly reassembled so that you interact with different participants.

The market has \texttt{60} identical customers. In every round of a period, each customer wishes to buy \textbf{one unit} of the product at the lowest possible price. Each customer is willing to pay at most \textbf{\texttt{4}} for this unit of the product.

All firms decide anew and \textbf{simultaneously} in every round for how many talers they want to sell their product. You can sell your product for a price of \texttt{1}, 1 taler, … or \texttt{5} (whole talers only). Your profit is the price multiplied by the number of units sold.
\begin{center}
\textbf{Profit = Price $\times$ Units sold}
\end{center}

The firm with the \textbf{lowest price in the respective round} sells its products as long as the price is not greater than \texttt{4}. Firms with a higher price do not sell their product. \textbf{The lowest price is the market price in the respective round.} Firms with a \textbf{price higher than the market price do not sell their products in that round.} If both firms offer their products at the same market price, the demand is divided equally between the two firms.

\paragraph{Example 1:} Firm A sets a price of 3 talers, Firm B sets a price of 3 talers. Firm A and B thus jointly set the lowest price. They each sell the same number of products (30 each) and therefore obtain the same profit of 90 talers.
\begin{center}
\begin{tabular}{lcc}
& \textbf{Firm A} & \textbf{Firm B} \\
\hline
Prices & 3 & 3 \\
Profits & 90 & 90 \\
\end{tabular}
\end{center}

\paragraph{Example 2:} Firm A sets a price of 1 taler, Firm B sets a price of 2 talers. Firm A sells the product to all 60 customers, Firm B sells nothing.
\begin{center}
\begin{tabular}{lcc}
& \textbf{Firm A} & \textbf{Firm B} \\
\hline
Prices & 1 & 2 \\
Profits & 60 & 0 \\
\end{tabular}
\end{center}

\paragraph{Example 3:} Firm A sets a price of 5 talers, Firm B sets a price of 5 talers. Customers are only willing to pay 4 talers. No firm sells a product.
\begin{center}
\begin{tabular}{lcc}
& \textbf{Firm A} & \textbf{Firm B} \\
\hline
Prices & 5 & 5 \\
Profits & 0 & 0 \\
\end{tabular}
\end{center}

\ \\
\textbf{Delegating to an Algorithm} \\

In this experiment, you can set the price for your firm's product in two ways: first, you can set the price yourself.\\

\textit{Only in the Recommendation treatment:}

Alternatively, you and the other participants have the option to \textbf{receive a price recommendation from an algorithm.} This decision is made \textbf{at the beginning of each period} and applies to all rounds of that period.

\textbf{Important:} Even if you decide to use algorithmic price setting, you \textbf{can override it at any time.} The algorithm's recommendation is therefore not binding and you retain full control. \\ \\

\textit{Only in the Outsourcing treatment:}

Alternatively, you and the other participants have the option to \textbf{let an algorithm set the price for you.} This decision is made \textbf{at the beginning of each period} and applies to all rounds of that period.

\textbf{Important:} If you choose the algorithm, it \textbf{takes over all price decisions for you.}\\


\textbf{If you choose the algorithm, you will still receive the profit your firm earns.} Using the algorithm is \textbf{free of charge} and aims to \textbf{maximize your profits throughout the entire period.}
\textbf{You decide before each period whether you want to use the algorithm.} Below we explain how the algorithm works. \\


\textbf{Description of the Algorithm}\\

In this experiment, a \textbf{pricing algorithm} is available to you that has been \textbf{tested in simulated scenarios}.

The algorithm uses \textbf{the prices from the previous round in your market} to determine a suitable price for each round. The algorithm has already learned in simulated scenarios and its behavior will not change during the experiment.

This version of the algorithm was selected for its performance in two key areas: \textbf{maximizing your profit} and \textbf{adaptability to different market strategies.}

\begin{itemize}
\item It is designed to set prices that, on average, maximize your profits across all rounds of a period.
\item It is designed to respond to different strategies of competitors and not be exploitable.
\end{itemize}

The decision to use this algorithm is yours. The algorithm is available to you in every period. \textbf{You can decide anew before the start of each period whether you want to use it or not.}

\subsection*{Instructions - German}

In diesem Experiment treffen Sie wiederholt Entscheidungen. Diese ermöglichen es Ihnen, echtes Geld zu verdienen. Wie viel Sie verdienen, ist abhängig von Ihren Entscheidungen und denen der anderen Teilnehmerinnen und Teilnehmer. \textbf{Unabhängig davon erhalten Sie \texttt{6} Euro für die Teilnahme.} \\

Im Experiment verwenden wir eine fiktive Geldeinheit namens Taler. Im Anschluss an das Experiment werden die Taler in Euro umgerechnet und Ihnen ausgezahlt. \textbf{Dabei entsprechen \texttt{140} Taler einem Euro.} \\

In diesem Experiment repräsentieren Sie eine Firma in einem virtuellen Produktmarkt. In jedem Markt verkauft eine weitere Firma das gleiche Produkt wie Sie. Diese Firma wird von einer anderen Experimentteilnehmerin bzw. einem Experimentteilnehmer repräsentiert. Alle Firmen bieten \texttt{60} Einheiten des vergleichbaren Produktes an. Bei der Produktion des Produktes entstehen für die Firmen keine Kosten. Das Experiment hat mehrere Durchgänge. Ein Durchgang hat mehrere Runden, wobei die genaue Anzahl durch einen Zufallsmechanismus entschieden wird. Sie spielen das Spiel für fünf Durchgänge. Sie treffen in jeder Runde eines Durchgangs auf dieselbe Firma (also eine Experimentteilnehmerin bzw. einen Experimentteilnehmer). Nach jedem Durchgang werden Sie zufällig erneut mit einer Firma gepaart.

\paragraph{Runde:} Eine Runde ist ein Teil innerhalb eines Durchgangs, in dem Sie mit der gleichen Teilnehmerin bzw. dem gleichen Teilnehmer interagieren. In jeder Runde treffen Sie Preisentscheidungen und erhalten Informationen zu Ihrem Gewinn und den Preisen der anderen Firmen.

\paragraph{Durchgang:} Ein Durchgang besteht aus mehreren Runden und stellt eine voll\-ständige Wiederholung des Spiels dar. Vor dem Beginn eines neuen Durchgangs wird der Markt zufällig neu zusammengestellt, sodass Sie mit unterschiedlichen Teilnehmerinnen/Teilnehmern interagieren.

Der Markt hat \texttt{60} identische Kunden. Jeder Kunde möchte in jeder Runde eines Durchgangs \textbf{eine Einheit} des Produkts möglichst günstig kaufen. Jeder Kunde ist bereit, maximal \textbf{\texttt{4}} für diese Einheit des Produkts auszugeben.

Alle Firmen entscheiden in jeder Runde erneut \textbf{gleichzeitig}, für wie viele Taler sie ihr Produkt verkaufen möchten. Sie können Ihr Produkt für einen Preis von \texttt{1}, \texttt{1} Taler, … oder \texttt{5} verkaufen (nur ganze Taler). Ihr Gewinn ist der Preis multipliziert mit der Anzahl an verkauften Einheiten.
\begin{center}
    \textbf{Gewinn = Preis $\times$ verkaufte Einheiten}
\end{center}

Die Firma mit dem \textbf{niedrigsten Preis in der jeweiligen Runde} verkauft ihre Produkte, solange der Preis nicht größer als \texttt{4} ist. Firmen mit einem höheren Preis verkaufen ihr Produkt nicht. \textbf{Der niedrigste Preis ist der Marktpreis in der jeweiligen Runde.} Firmen mit einem \textbf{Preis, der größer als der Marktpreis ist, verkaufen ihre Produkte in der jeweiligen Runde nicht.} Sollten beide Firmen ihre Produkte für den gleichen Marktpreis verkaufen wollen, teilt sich die Nachfrage gleichmäßig auf die zwei Firmen auf.

\paragraph{Beispiel 1:} Firma A setzt einen Preis von 3 Taler, Firma B setzt einen Preis von 3 Taler. Somit haben Firma A und B gemeinsam den niedrigsten Preis gesetzt. Sie verkaufen beide die gleiche Anzahl an Produkten (jeweils 30) und bekommen damit den gleichen Gewinn von 90 Talern.
\begin{center}
\begin{tabular}{lcc}
 & \textbf{Firma A} & \textbf{Firma B} \\
\hline
Preise & 3 & 3 \\
Gewinne & 90 & 90 \\
\end{tabular}
\end{center}

\paragraph{Beispiel 2:} Firma A setzt einen Preis von 1 Taler, Firma B setzt einen Preis von 2 Taler. Firma A verkauft das Produkt an alle 60 Kunden, Firma B verkauft nichts.
\begin{center}
\begin{tabular}{lcc}
 & \textbf{Firma A} & \textbf{Firma B} \\
\hline
Preise & 1 & 2 \\
Gewinne & 60 & 0 \\
\end{tabular}
\end{center}

\paragraph{Beispiel 3:} Firma A setzt einen Preis von 5 Taler, Firma B setzt einen Preis von 5 Taler. Kunden sind nur bereit 4 Taler zu zahlen. Keine Firma verkauft ein Produkt.
\begin{center}
\begin{tabular}{lcc}
 & \textbf{Firma A} & \textbf{Firma B} \\
\hline
Preise & 5 & 5 \\
Gewinne & 0 & 0 \\
\end{tabular}
\end{center}

\ \\
\textbf{Delegieren an einen Algorithmus} \\

In diesem Experiment können Sie den Preis für das Produkt Ihrer Firma auf zwei Arten festlegen: Zum einen können Sie selbst den Preis setzen.\\

\textit{Nur im Recommendation Treatment:}

Alternativ haben Sie und die anderen Teilnehmerin und Teilnehmer die Möglichkeit, \textbf{eine Preisempfehlung von einem Algorithmus zu erhalten.} Diese Entscheidung wird \textbf{zu Beginn jedes Durchgangs} getroffen und gilt für alle Runden dieses Durchgangs.

\textbf{Wichtig ist:} Auch wenn Sie sich entscheiden, die algorithmische Preissetzung zu nutzen, so \textbf{können Sie diese jederzeit überschreiben.} Die Empfehlung des Algorithmus ist also nicht binden und Sie behalten die volle Kontrolle. \\

\textit{Nur im Outsourcing Treatment:}

Alternativ haben Sie und die anderen Teilnehmerinnen und Teilnehmer die Möglichkeit, \textbf{einen Algorithmus den Preis für Sie festlegen zu lassen.} Diese Entscheidung wird \textbf{zu Beginn jedes Durchgangs} getroffen und gilt für alle Runden dieser Durchgangs.

\textbf{Wichtig ist:} Wenn Sie sich für den Algorithmus entscheiden, \textbf{übernimmt dieser alle Preisentscheidungen für Sie.}\\


\textbf{Falls Sie sich für den Algorithmus entscheiden, erhalten Sie weiterhin den Gewinn, den Ihre Firma erzielt.} Die Verwendung des Algorithmus ist \textbf{kostenlos} und zielt darauf ab, \textbf{Ihre Gewinne über den gesamten Durchgang zu maximieren.}\
\textbf{Sie treffen vor jedem Durchgang erneut die Entscheidung, ob Sie den Algorithmus nutzen wollen.} Im Folgenden erklären wir Ihnen, wie der Algorithmus funktioniert. \\


\textbf{Beschreibung des Algorithmus}\\

In diesem Experiment steht Ihnen ein \textbf{Preisalgorithmus} zur Verfügung, der \textbf{in simulierten Szenarien getestet} wurde.

Der Algorithmus verwendet \textbf{die Preise aus der vorherigen Runde aus Ihrem Markt}, um einen geeigneten Preis für jede Runde zu ermitteln. Der Algorithmus hat bereits in simulierten Szenarien gelernt und das Verhalten des Algorithmus wird sich im Experiment nicht ändern.

Diese Version des Algorithmus wurde aufgrund seiner Leistung in zwei Schlüsselbereichen ausgewählt: \textbf{Maximierung Ihres Gewinns} und \textbf{Anpassungsfähigkeit an verschiedene Marktstrategien.}

\begin{itemize}
\item Er ist darauf ausgelegt, Preise festzulegen, die im Durchschnitt Ihre Gewinne über alle Runden eines Durchgangs maximieren.
\item Er ist darauf ausgelegt, auf unterschiedliche Strategien von Wettbewerbern reagieren zu können und nicht ausnutzbar zu sein.
\end{itemize}

Die Entscheidung, diesen Algorithmus zu verwenden, liegt bei Ihnen. Der Algorithmus steht Ihnen in jedem Durchgang zur Verfügung. \textbf{Sie können vor dem Beginn jedes Durchgangs neu entscheiden, ob Sie ihn verwenden möchten oder nicht.}
}
\onlineappendixsection{Experimental design}

\subsection*{Experimental design - English translation}

\begin{figure}[H]
    \centering
    \includegraphics[width=1\linewidth, trim=20cm 5cm 20cm 3cm, clip]{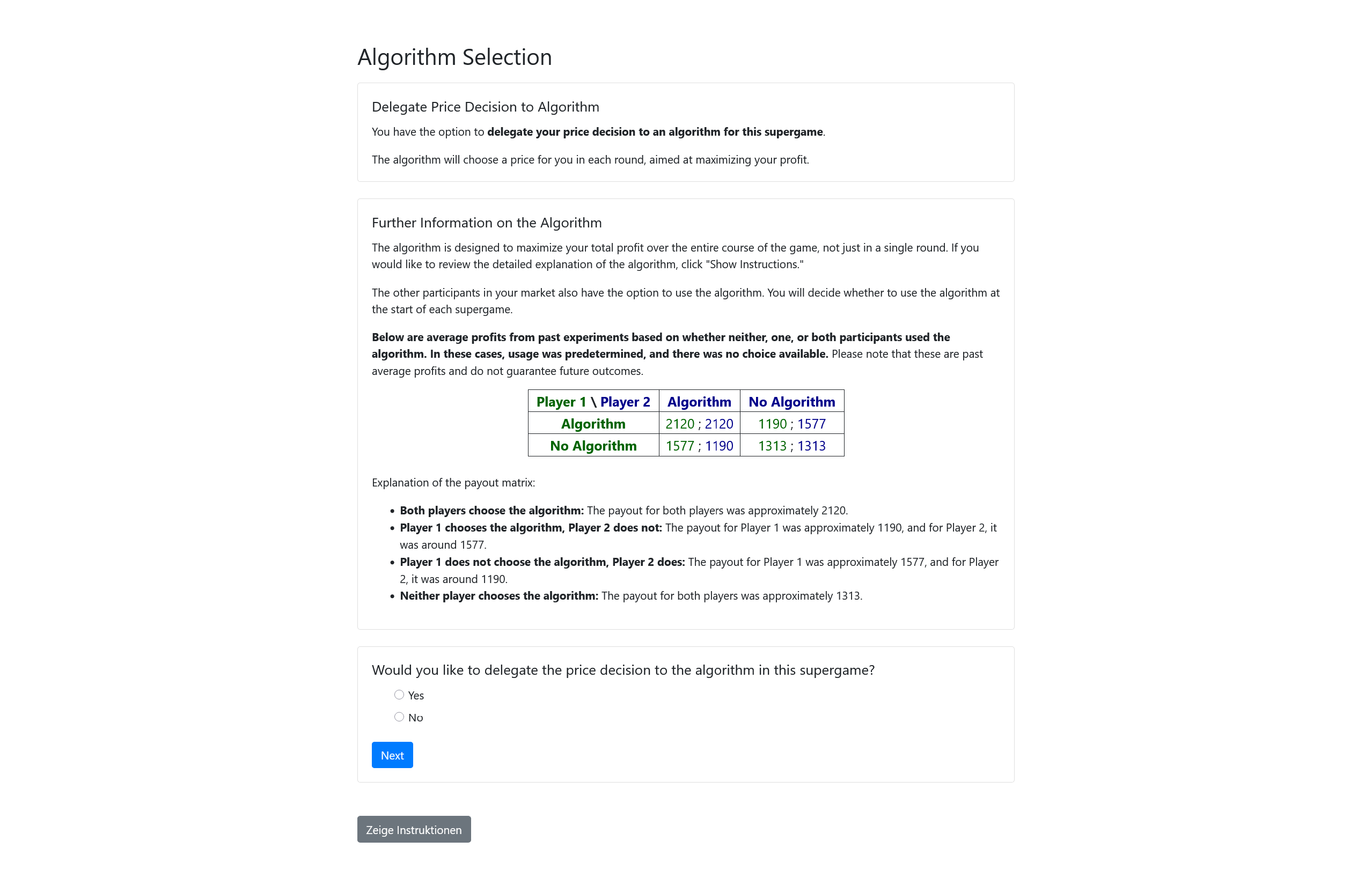}
    \caption{Decision screen for algorithm adoption in \textsc{Outsourcing}}
    \label{fig:adoption_screen_out}
\end{figure}

\begin{figure}[H]
    \centering
    \includegraphics[width=1\linewidth, trim=20cm 5cm 20cm 3cm, clip]{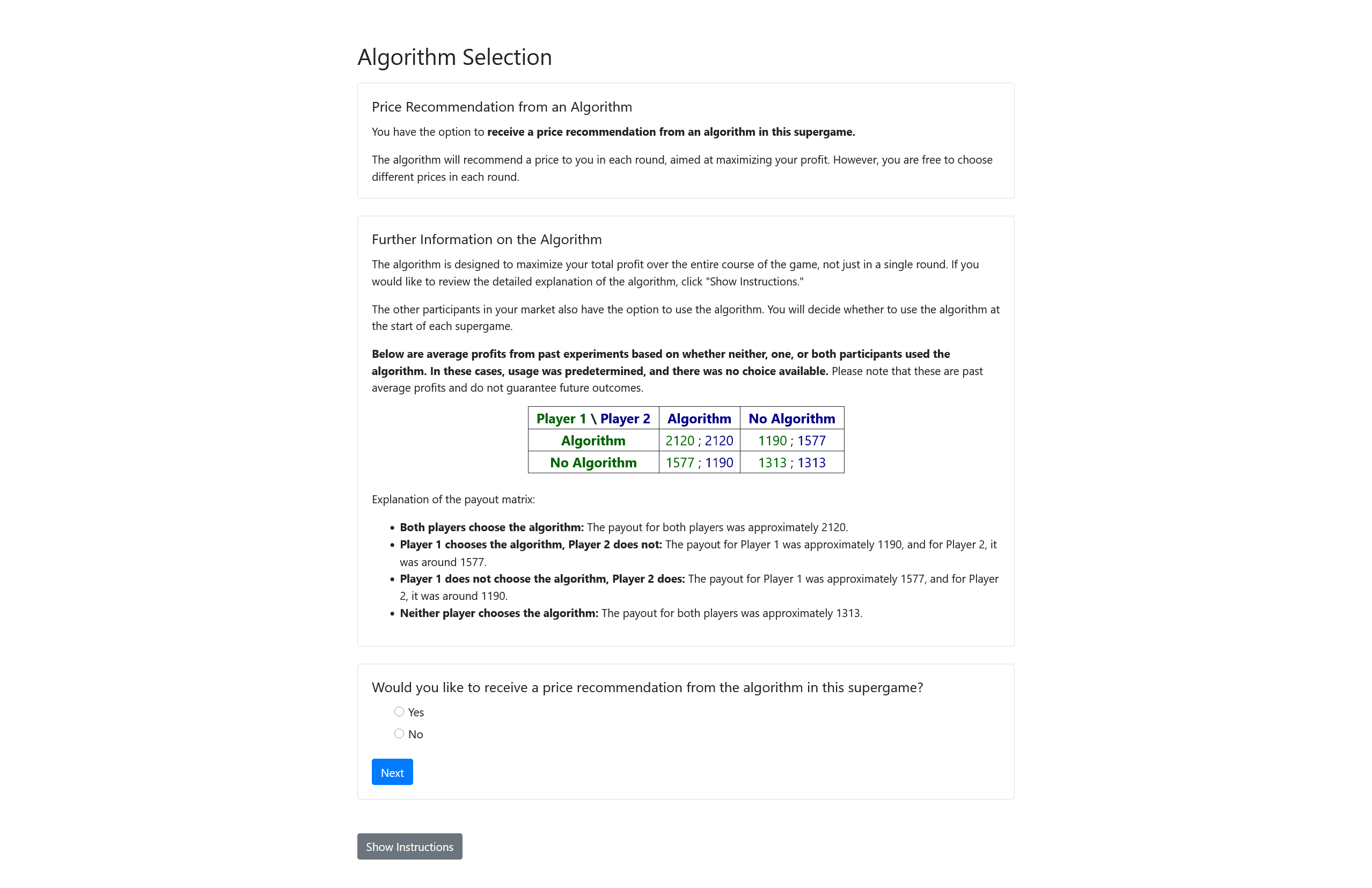}
    \caption{Decision screen for algorithm adoption in \textsc{Recommendation}}
    \label{fig:adoption_screen_rec}
\end{figure}

\begin{figure}[H]
    \centering
    \includegraphics[width=1\linewidth, trim=20cm 24cm 20cm 3cm, clip]{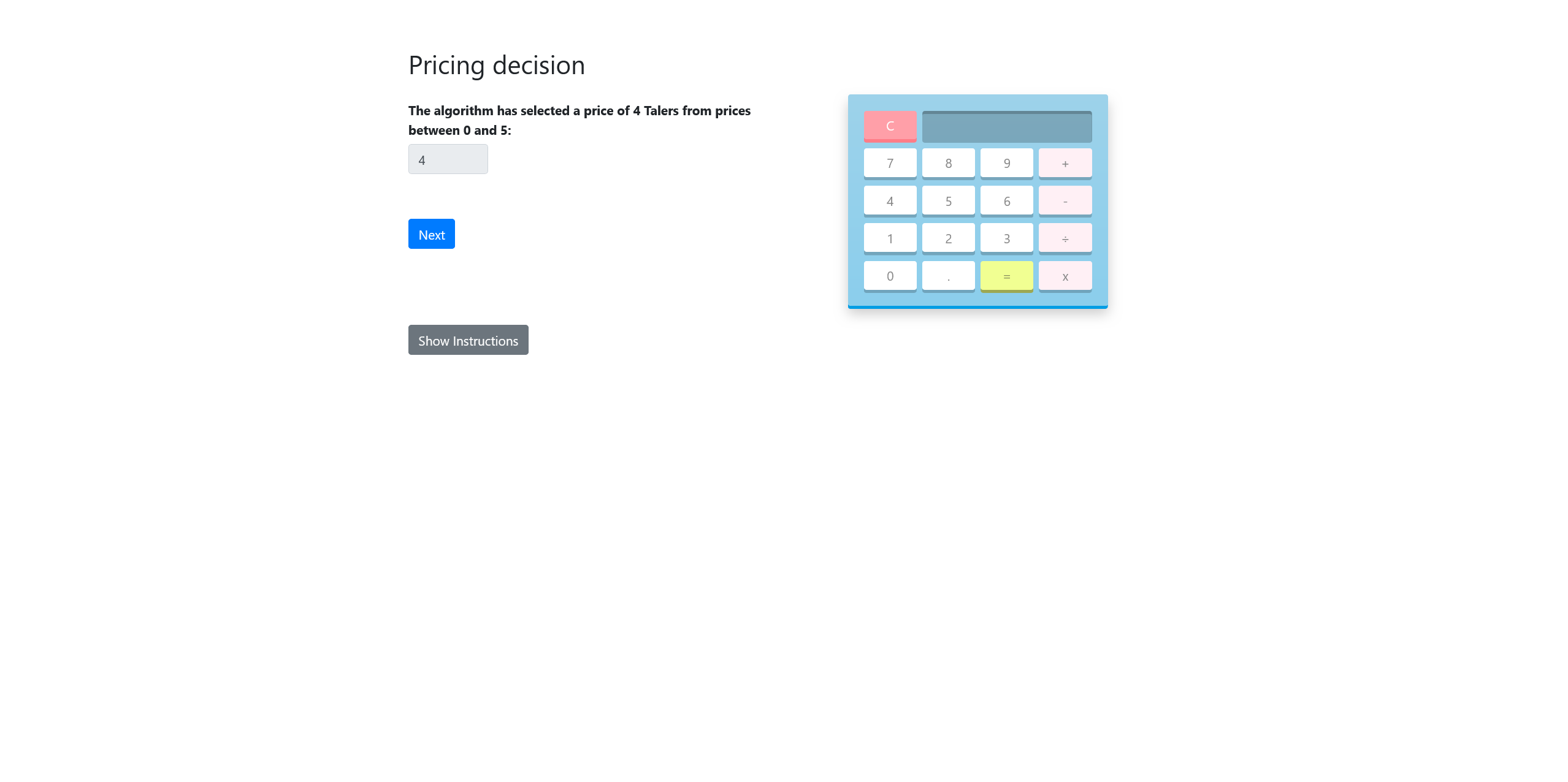}
    \caption{Decision screen in \textsc{Outsourcing} with algorithm adoption}
    \label{fig:decision_screen_out}
\end{figure}

\begin{figure}[H]
        \centering
        \includegraphics[width=1\linewidth, trim=20cm 24cm 20cm 3cm, clip]{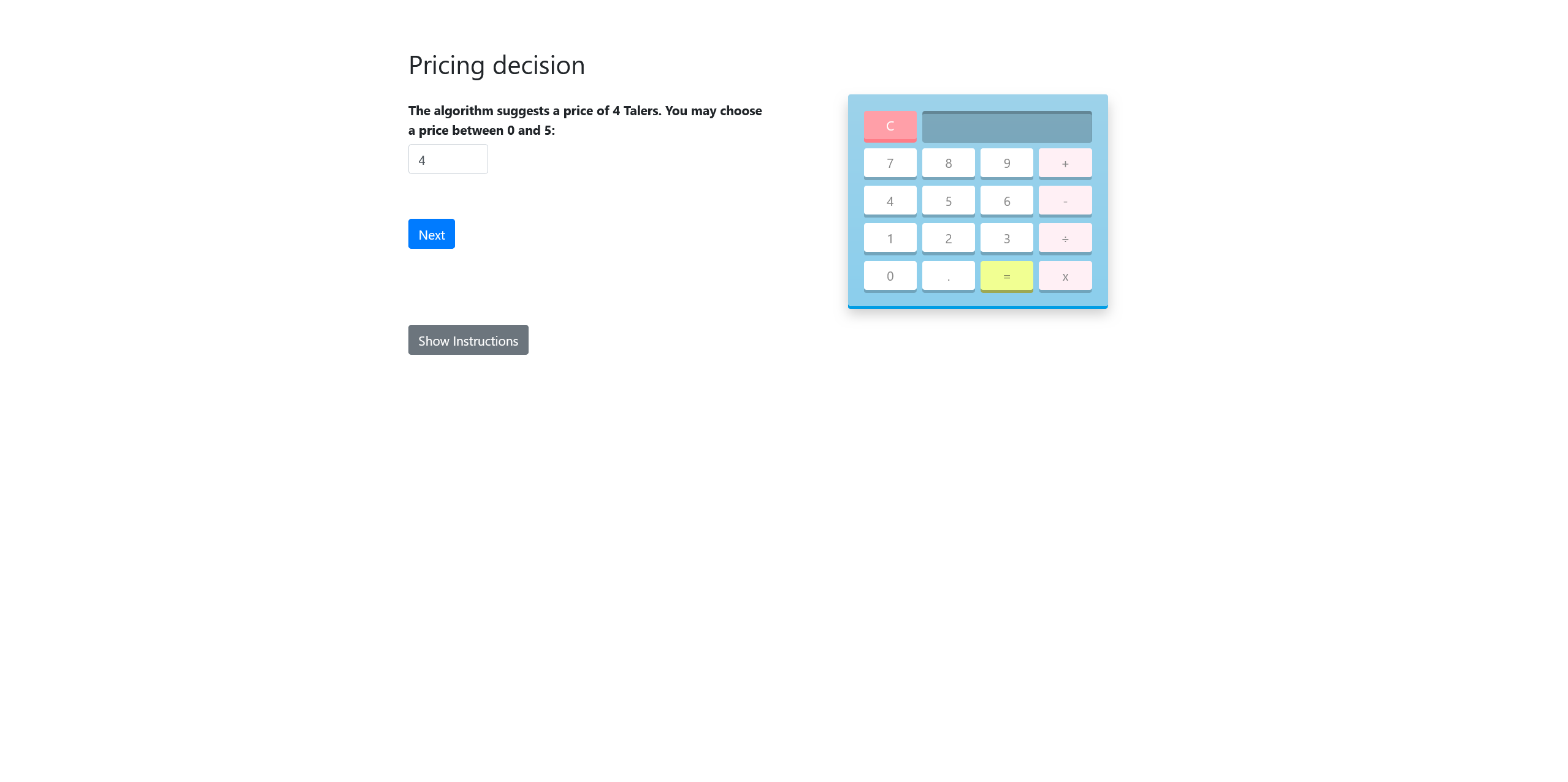}
        \caption{Decision screen in \textsc{Recommendation} with algorithm adoption}
        \label{fig:decision_screen_rec}
\end{figure}

\begin{figure}[H]
        \centering
        \includegraphics[width=1\linewidth, trim=20cm 17cm 20cm 3cm, clip]{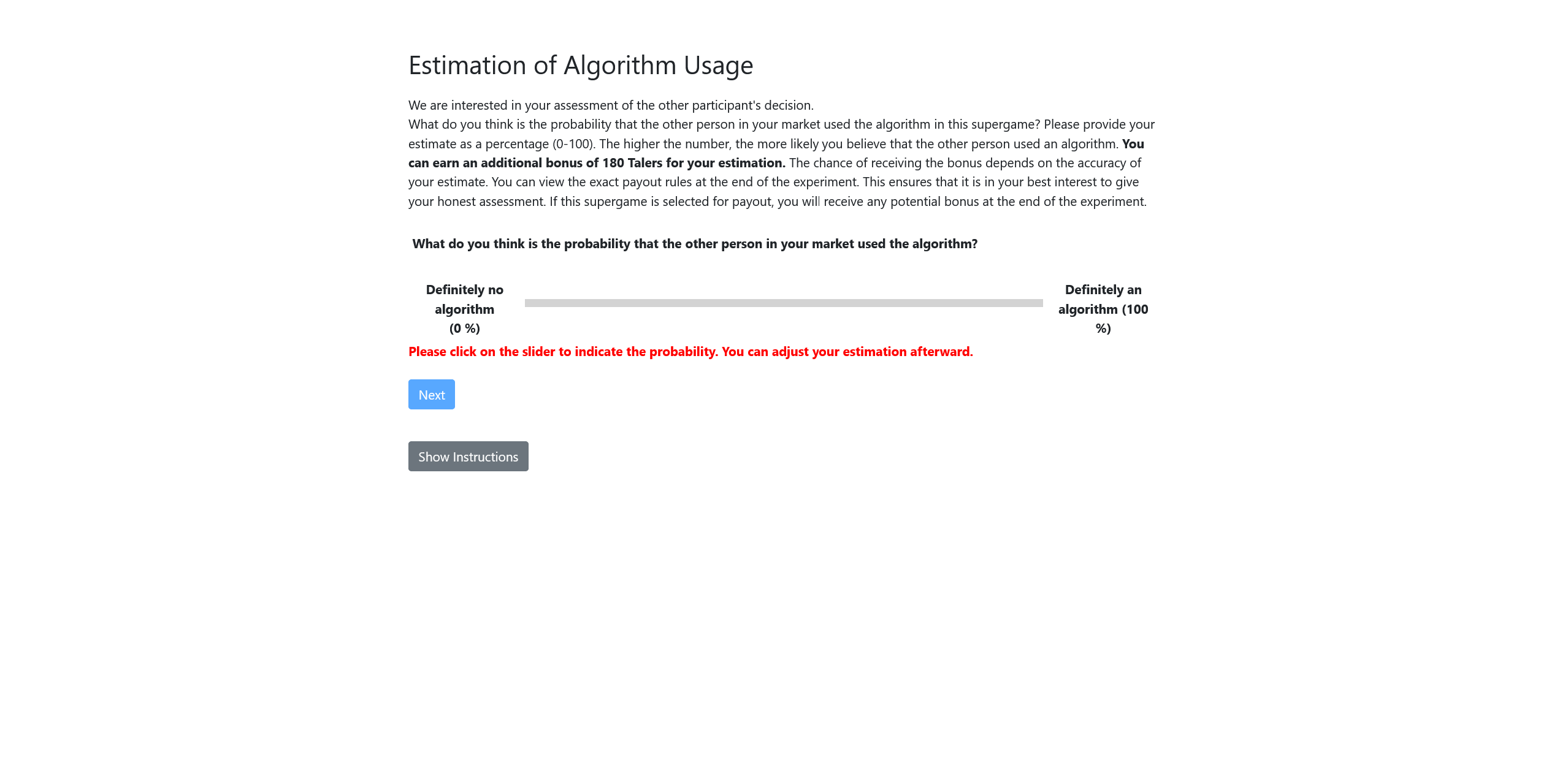}
        \caption{Belief elicitation screen in algorithmic treatments}
        \label{fig:elicitation_screen}
\end{figure}

\subsection*{Experimental Design - German}

\begin{figure}[H]
    \centering
    \includegraphics[width=1\linewidth, trim=15cm 36.5cm 15cm 2cm, clip]{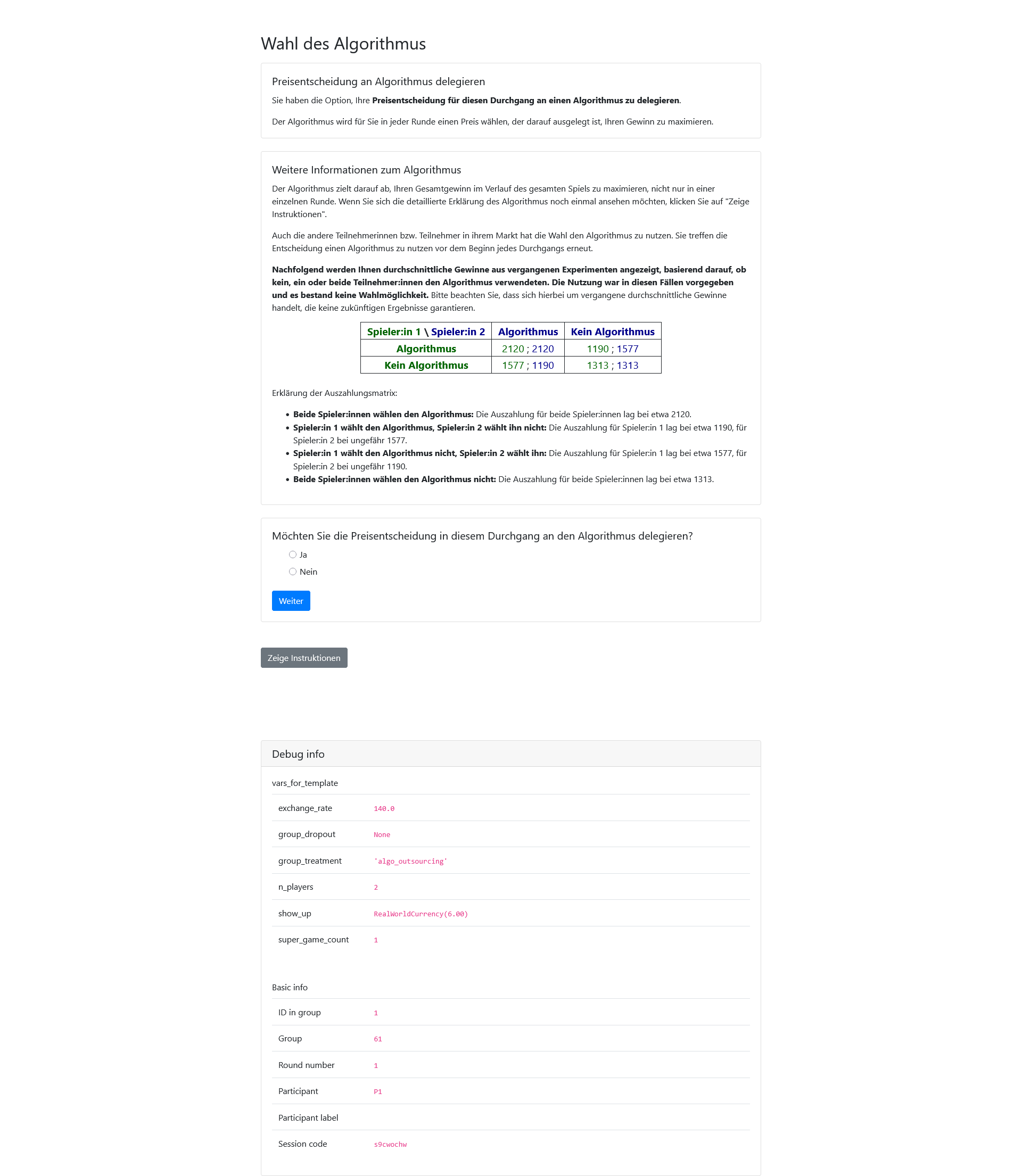}
    \caption{Decision screen for algorithm adoption in \textsc{Outsourcing}}
    \label{fig:adoption_screen_out_de}
\end{figure}

\begin{figure}[H]
    \centering
    \includegraphics[width=1\linewidth, trim=15cm 36.5cm 15cm 2cm, clip]{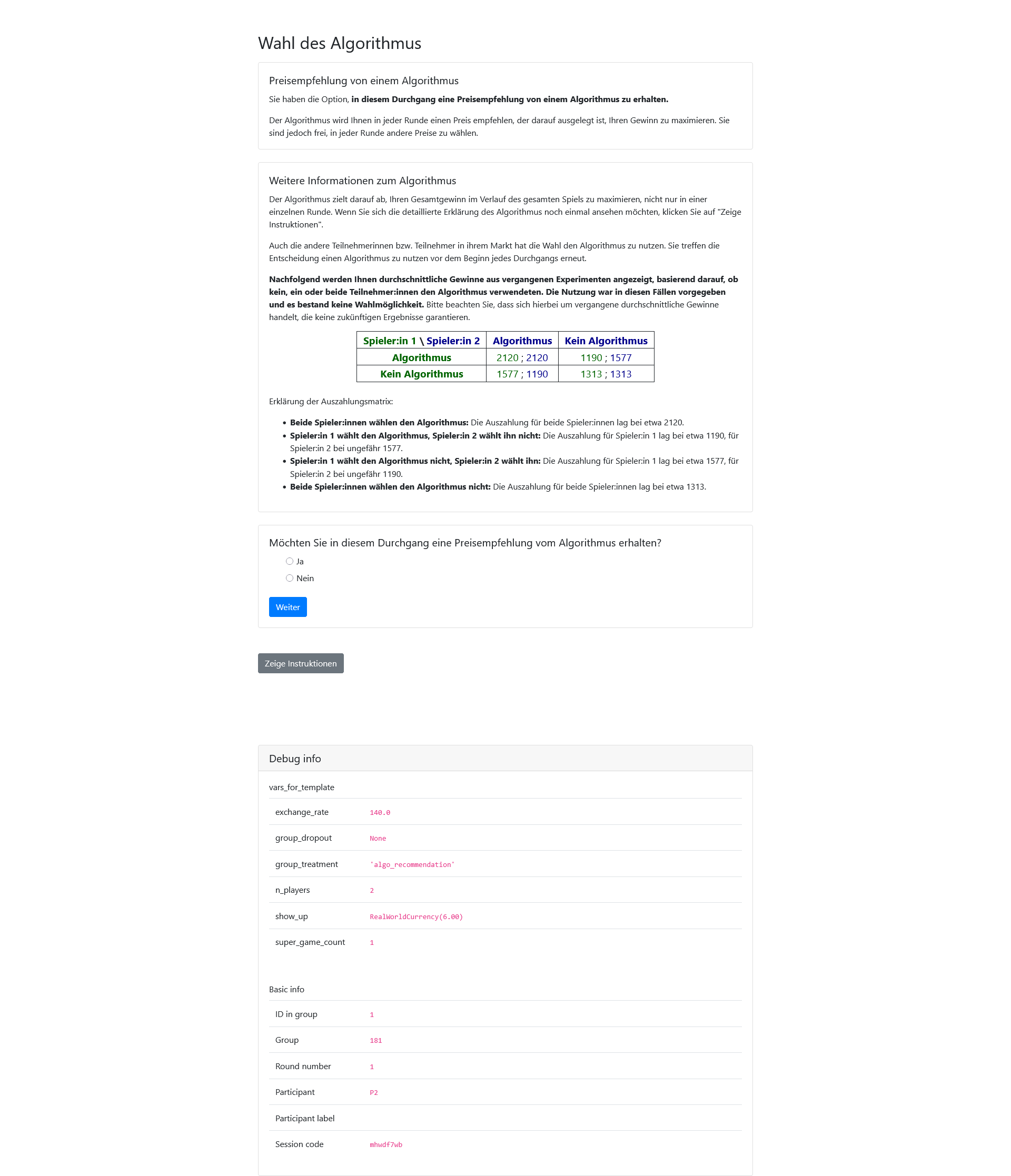}
    \caption{Decision screen for algorithm adoption in \textsc{Recommendation}}
    \label{fig:adoption_screen_rec_de}
\end{figure}

\begin{figure}[H]
    \centering
    \includegraphics[width=1\linewidth, trim=15cm 36.5cm 15cm 2cm, clip]{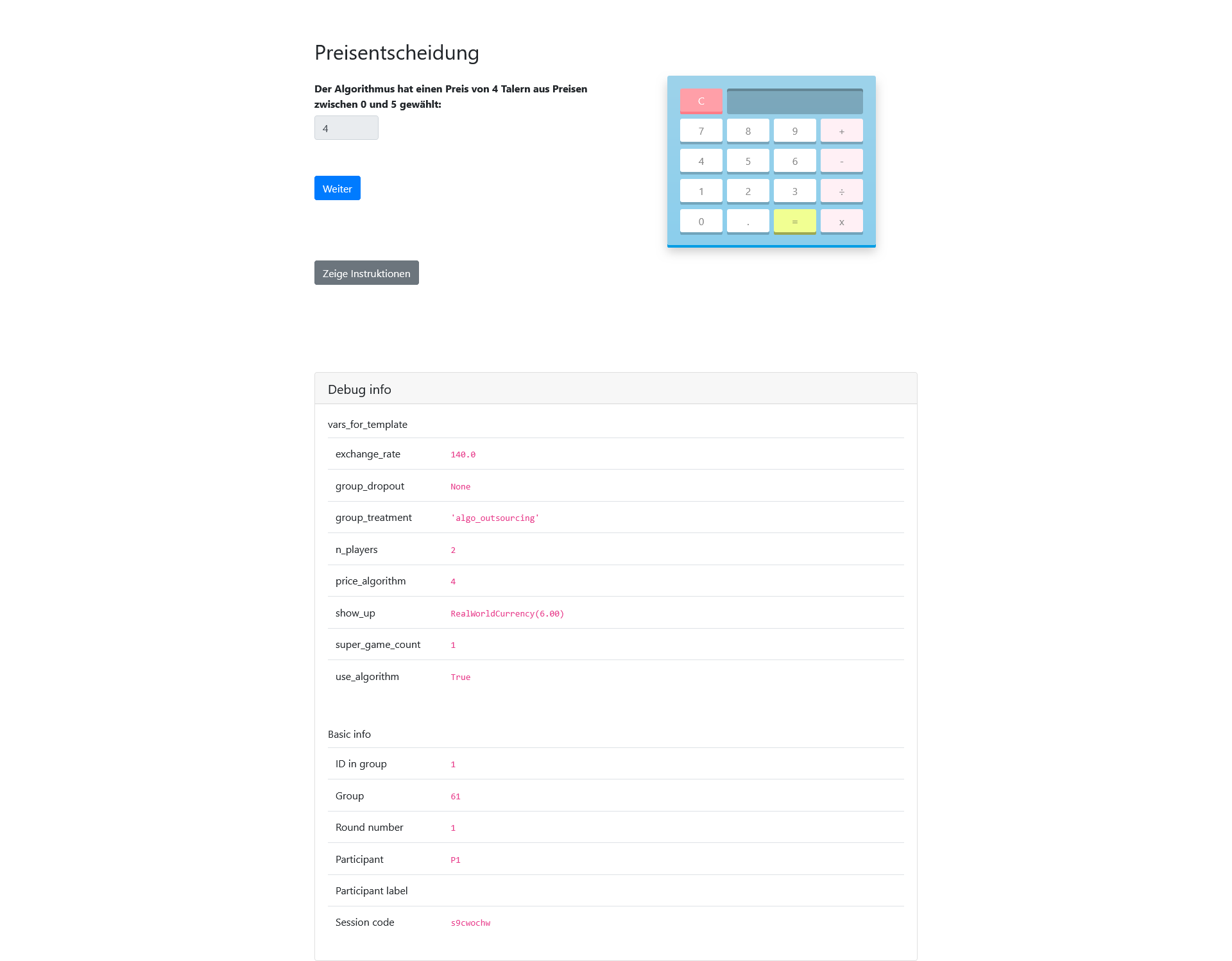}
    \caption{Decision screen in \textsc{Outsourcing} with algorithm adoption}
    \label{fig:decision_screen_out_de}
\end{figure}

\begin{figure}[H]
        \centering
        \includegraphics[width=1\linewidth, trim=15cm 36.5cm 15cm 2cm, clip]{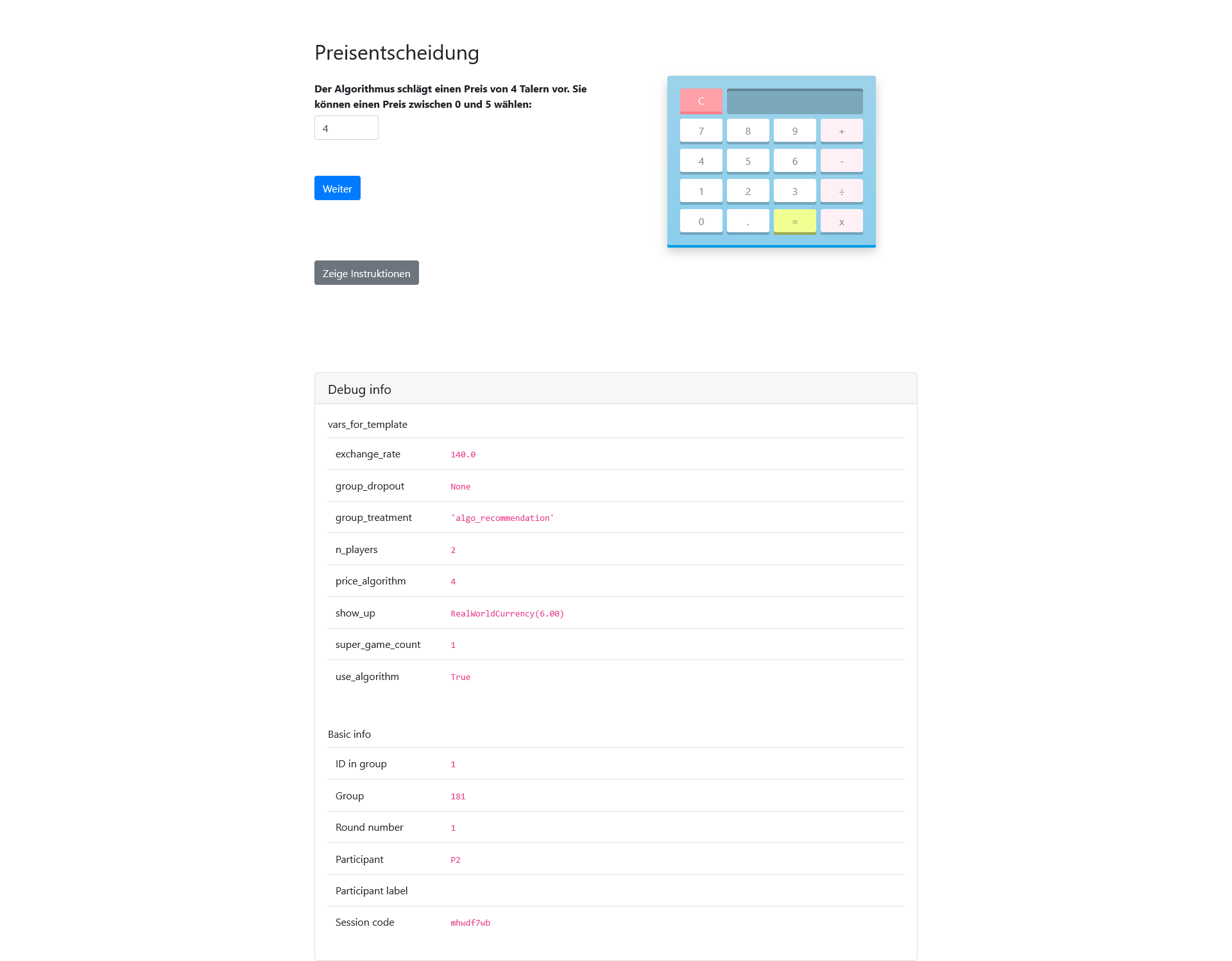}
        \caption{Decision screen in \textsc{Recommendation} with algorithm adoption}
        \label{fig:decision_screen_rec_de}
\end{figure}

\begin{figure}[H]
        \centering
        \includegraphics[width=1\linewidth, trim=15cm 31.5cm 15cm 2cm, clip]{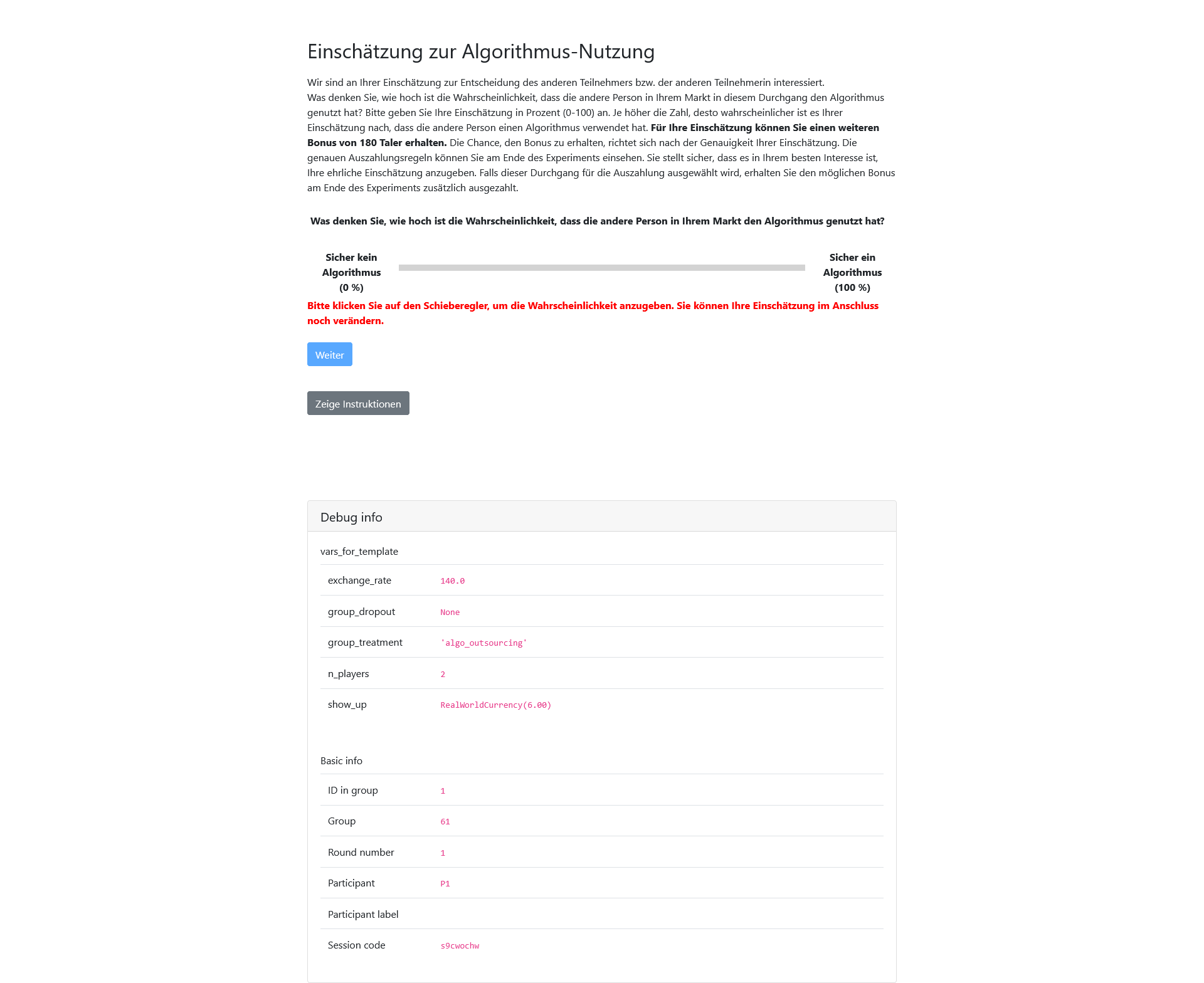}
        \caption{Belief elicitation screen in algorithmic treatments}
        \label{fig:elicitation_screen_de}
\end{figure}

\onlineappendixsection{Results section}

\begin{table}[H]
\centering
\caption{Sample composition across treatments: Means or proportions and test $p$-values}\label{tab:sample_composition_test}
\begin{threeparttable}
\begin{tabular}{llcccr}
\toprule
\multicolumn{2}{l}{Statistic and Level} & \multicolumn{1}{c}{Baseline} & \multicolumn{1}{c}{Outsourcing} & \multicolumn{1}{c}{Recommendation} & \multicolumn{1}{c}{$p$-value} \\
\midrule
\multicolumn{6}{l}{\textsc{Age and Experiment Participation}} \\
\quad Age & & 25.24 & 23.61 & 23.85 & 0.9772 \\
\quad Number of Experiments & & 5.41 & 5.03 & 4.12 & 0.3075 \\
\addlinespace
\multicolumn{5}{l}{\textsc{Gender}} & 0.2739 \\
\quad Diverse & & 0.00 & 0.01 & 0.02 &  \\
\quad No Answer & & 0.02 & 0.00 & 0.01 &  \\
\quad Male & & 0.43 & 0.42 & 0.32 &  \\
\quad Female & & 0.55 & 0.57 & 0.65 &  \\
\addlinespace
\multicolumn{5}{l}{\textsc{Educational Attainment}} & 0.6061\\
\quad Secondary School (Abitur) & & 0.65 & 0.65 & 0.74 &  \\
\quad Bachelor's Degree & & 0.24 & 0.22 & 0.18 &  \\
\quad Vocational Qualification & & 0.01 & 0.03 & 0.02 &  \\
\quad Master's Degree & & 0.08 & 0.10 & 0.06 &  \\
\quad PhD & & 0.02 & 0.00 & 0.00 &  \\
\bottomrule
\end{tabular}
\begin{tablenotes}
\item \scriptsize Values report means (for continuous variables: age and number of experiments) or column proportions (for categorical variables: gender, education level) by treatment group. The p-values are computed using Kruskal-Wallis rank-sum tests for continuous variables and Pearson’s chi-squared tests for categorical variables (or Fisher's exact test where appropriate). Each observation corresponds to a unique participant.
\end{tablenotes}
\end{threeparttable}
\end{table}

\begin{table}[H]
\centering
\caption{Non-parametric tests: Adoption and average market prices}
\label{tab:np_mwu_main}
\begin{threeparttable}
\begin{tabular}{llcccc}
\toprule
\multicolumn{2}{l}{Statistic and Level} &
\multicolumn{1}{c}{Baseline} &
\multicolumn{1}{c}{Outsourcing} &
\multicolumn{1}{c}{Recommendation} &
\multicolumn{1}{c}{$p$-value} \\
\midrule
\multicolumn{6}{l}{\textsc{Algorithm Adoption \,(MWU; no Baseline)}} \\
\quad ALL  (O\,vs\,R)      & & -- & 0.53 & 0.83 & 0.0013 \\
\quad SG1  (O\,vs\,R)      & & -- & 0.70 & 0.86 & 0.0232 \\
\quad SG5  (O\,vs\,R)      & & -- & 0.47 & 0.77 & 0.0010 \\
\addlinespace
\multicolumn{6}{l}{\textsc{Average Market Prices \,(MWU)}} \\
\quad ALL  (O\,vs\,R)      & & 2.95 & 2.79 & 2.40 & 0.1212 \\
\quad ALL  (B\,vs\,O)      & & 2.95 & 2.79 & 2.40 & 0.1859 \\
\quad ALL  (B\,vs\,R)      & & 2.95 & 2.79 & 2.40 & 0.0757 \\[2pt]
\quad SG1  (O\,vs\,R)      & & 2.52 & 3.16 & 2.09 & 0.0013 \\
\quad SG1  (B\,vs\,O)      & & 2.52 & 3.16 & 2.09 & 0.0140 \\
\quad SG1  (B\,vs\,R)      & & 2.52 & 3.16 & 2.09 & 0.0538 \\[2pt]
\quad SG5  (O\,vs\,R)      & & 3.12 & 2.72 & 2.22 & 0.0820 \\
\quad SG5  (B\,vs\,O)      & & 3.12 & 2.72 & 2.22 & 0.1123 \\
\quad SG5  (B\,vs\,R)      & & 3.12 & 2.72 & 2.22 & 0.0017 \\
\bottomrule
\end{tabular}
\begin{tablenotes}
\item \scriptsize Means and Mann–Whitney U (MWU) tests are computed at the matching-group level and compare treatment pairs.
\end{tablenotes}
\end{threeparttable}
\end{table}

\begin{table}[H]
\caption{Belief accuracy by Treatment $\times$ Opponent Adoption}
\begin{center}
\begin{tabular}{l c c}
\toprule
 & Belief Accuracy & $H_{0}:$ Belief = 0.5 \\
\midrule
Recommendation $\times$ Opponent No-Adopt & $0.54$  & $p=0.3510$ \\
                                          & $(0.04)$     \\
Recommendation $\times$ Opponent Adopt    & $0.59$  & $p=0.0045$ \\
                                          & $(0.03)$     \\
Outsourcing $\times$ Opponent No-Adopt    & $0.62$  & $p=0.0000$ \\
                                          & $(0.03)$     \\
Outsourcing $\times$ Opponent Adopt       & $0.81$  & $p=0.0000$ \\
                                          & $(0.03)$     \\
\midrule
Sub-sample & Out \& Rec \\
$\# $ observations & 1000 \\
$\# $ cluster & 20 \\
R$^2$ & 0.07 \\
\bottomrule
\multicolumn{3}{p{0.86\textwidth}}{\scriptsize{Results are based on linear regressions without an intercept, estimated with standard errors clustered at the matching-group level. Belief accuracy is defined as $1-\lvert b_{is}-d_{is}\rvert$, where $b_{is}\in[0,1]$ is the participant’s stated belief and $d_{is}\in\{0,1\}$ is the realized opponent-adoption indicator in a given supergame; the index equals 1 for a perfectly accurate belief and tends toward 0 as the belief diverges from the outcome. Each coefficient equals the subgroup mean. The second column reports two-sided $p$-values for $H_{0}\!:\,\mu = 0.5$ (not better than chance). Wald tests of treatment differences: Outsourcing $\times$ Opp.\ No--Adopt $-$ Recommendation $\times$ Opp.\ No--Adopt $=0$: $p<0.1^{*}$; Outsourcing $\times$ Opp.\ Adopt $-$ Recommendation $\times$ Opp.\ Adopt $=0$: $p<0.01^{***}$. }}
\end{tabular}\label{table:belief_result2}
\end{center}
\end{table}

\begin{table}[H]
\caption{Market price difference: Round 1 vs. subsequent rounds}
\label{tab:C04_first_round_effect_by_treatment}
\begin{center}
\begin{tabular}{l c c}
\toprule
& \multicolumn{2}{c}{Market Prices} \\
\cmidrule(lr){2-3}
& (1) & (2) \\
\midrule
\begin{tabular}{@{}l@{}}Round 1 $\times$ Baseline\end{tabular}       & $-0.11$      & $-0.11$      \\
                                                                     & $(0.12)$     & $(0.08)$     \\
\begin{tabular}{@{}l@{}}Round 1 $\times$ Outsourcing\end{tabular}    & $0.45^{***}$ & $0.55^{***}$ \\
                                                                     & $(0.08)$     & $(0.09)$     \\
\begin{tabular}{@{}l@{}}Round 1 $\times$ Recommendation\end{tabular} & $0.96^{***}$ & $0.86^{***}$ \\
                                                                     & $(0.10)$     & $(0.11)$     \\
\midrule
FE & Yes & Yes \\
$\# $ observations & 15300 & 15300 \\
$\# $ cluster & 30 & 30 \\
R$^2$ & 0.78 & 0.80 \\
\bottomrule
\multicolumn{3}{p{0.55\textwidth}}{\scriptsize{Results are from linear regressions estimated without an intercept. Dependent variable: market price. Standard errors are clustered at the matching-group level. All specifications include market and supergame fixed effects, so coefficients on $Round\; 1 \times treatment$ are identified from within-market, within-supergame variation in prices between the first and later rounds of the same treatment. Model (2) is weighted by supergame lengths. The differences in first-round effects among \textsc{Outsourcing}, \textsc{Recommendation}, and \textsc{Baseline} are highly significant in the unweighted specification (regression (1), $p<0.01$, Wald tests). In the weighted specification (regression (2)), the differences between \textsc{Outsourcing} and \textsc{Baseline} and between \textsc{Recommendation} and \textsc{Baseline} are highly significant ($p<0.01$, Wald tests), while the difference between \textsc{Outsourcing} and \textsc{Recommendation} is significant ($p<0.05$, Wald test). Significance levels: *** $p<0.01$; ** $p<0.05$; * $p<0.1$.}}
\end{tabular}
\end{center}
\end{table}

\begin{figure}[H]
        \centering
        \includegraphics[width=0.85\linewidth]{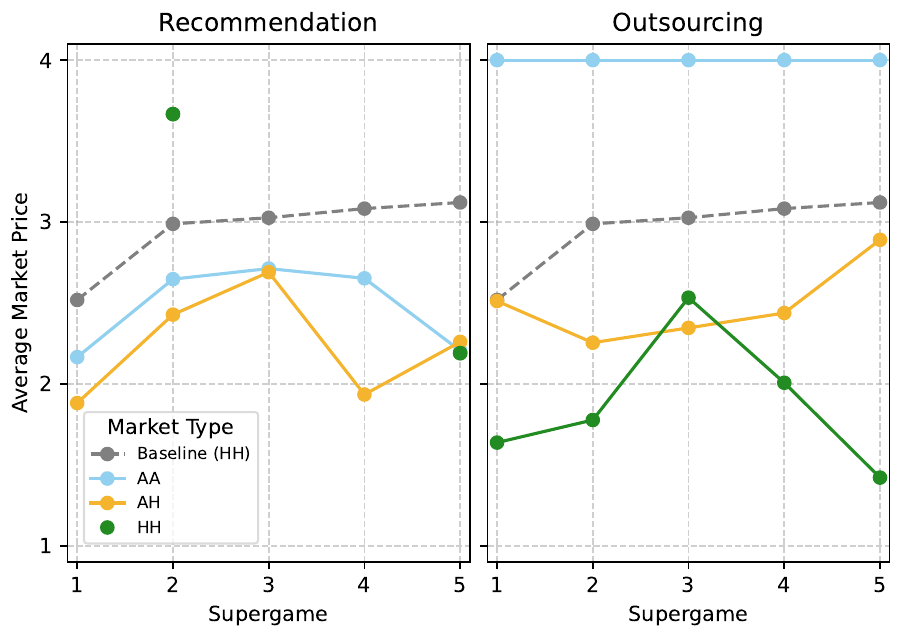}
        \caption{Average market price per market type and supergame}
        \label{fig:wprice_mkttype}
\end{figure}

\begin{figure}[H]
        \centering
        \includegraphics[width=0.85\linewidth]{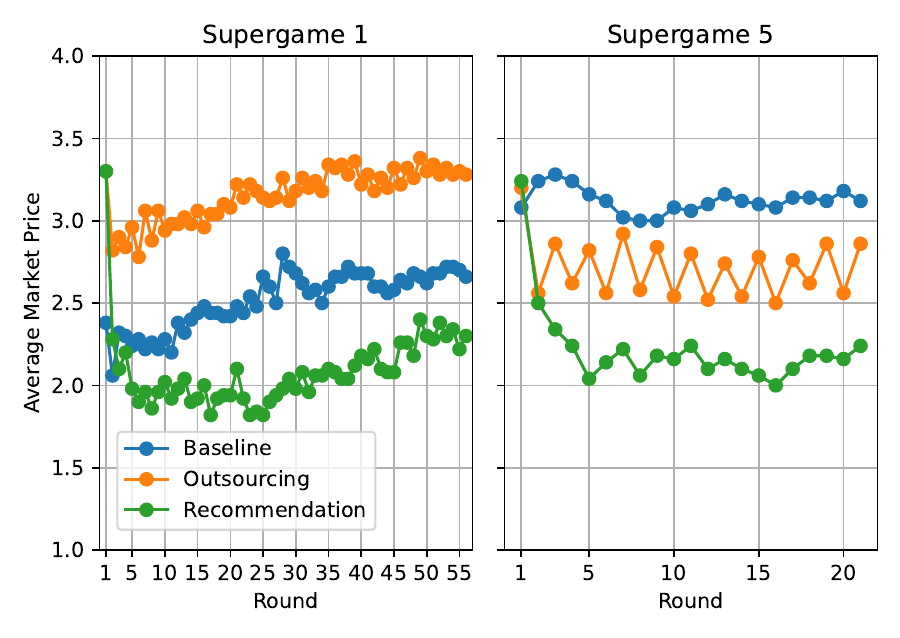}
        \caption{Average market prices by round and supergame}
        \label{fig:avgprices_round_sg1vs5}
\end{figure}

\begin{figure}[H]
        \centering
        \includegraphics[width=0.85\linewidth]{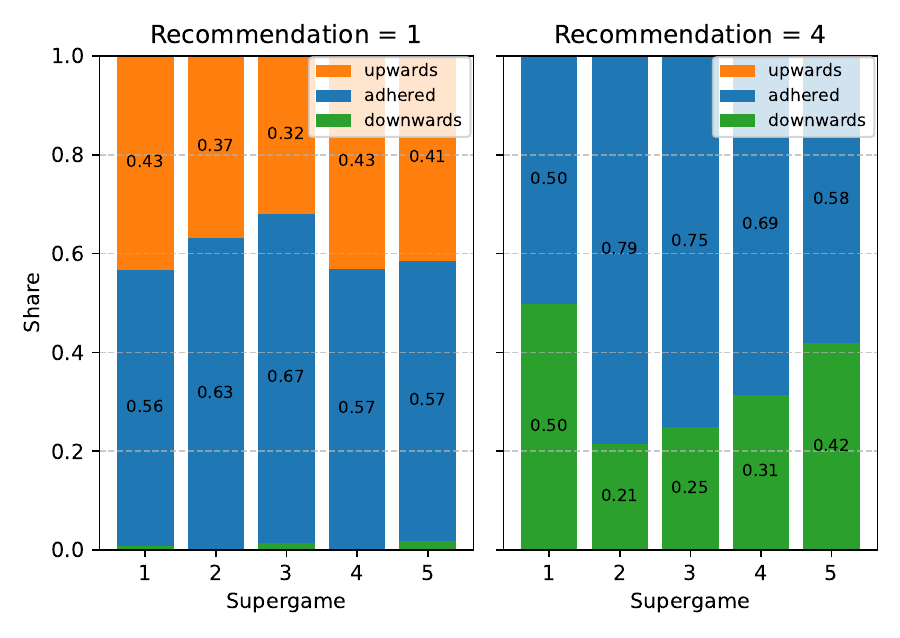}
        \caption{Adherence and deviations by type of algorithmic recommendations}
        \label{fig:deviation_shares_recommendation_1_4}
\end{figure}

\begin{figure}[H]
        \centering
        \includegraphics[width=0.85\linewidth]{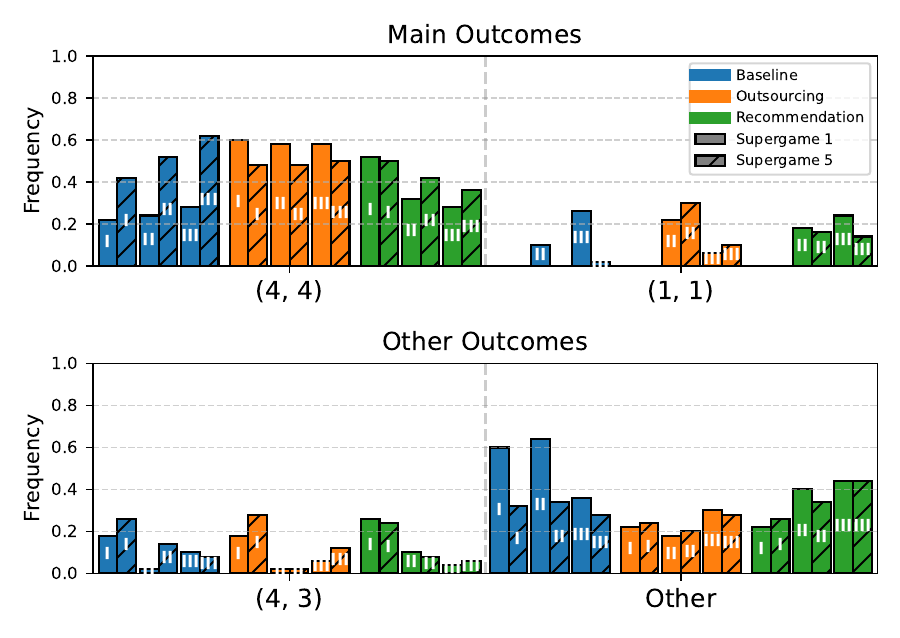}
        \caption{Outcome distribution in the first three rounds across treatments}
        \label{fig:outcome_distribution_first3_all}
\end{figure}

\begin{table}[H]
\caption{Per-period profits, in supergames 1 and 5 for \textsc{Outsourcing} and \textsc{Recommendation} treatments.}
\label{tab:normalized_profits_sg1_sg5}
\centering

\begin{minipage}{0.5\textwidth}
\centering
\begin{tabular}{lcc}
\multicolumn{3}{c}{\sc Outsourcing (SG1)} \\
\toprule
 & Algorithm & No Algorithm \\
\midrule
Algorithm    & (120.00, 120.00) & (54.74, 95.99) \\
No Algorithm & (95.99, 54.74)   & (49.15, 49.15) \\
\bottomrule
\end{tabular}
\end{minipage}

\bigskip

\begin{minipage}{0.5\textwidth}
\centering
\begin{tabular}{lcc}
\multicolumn{3}{c}{\sc Outsourcing (SG5)} \\
\toprule
 & Algorithm & No Algorithm \\
\midrule
Algorithm    & (120.00, 120.00) & (72.11, 101.30) \\
No Algorithm & (101.30, 72.11)   & (42.67, 42.67) \\
\bottomrule
\end{tabular}
\end{minipage}

\bigskip

\begin{minipage}{0.5\textwidth}
\centering
\begin{tabular}{lcc}
\multicolumn{3}{c}{\sc Recommendation (SG1)} \\
\toprule
 & Algorithm & No Algorithm \\
\midrule
Algorithm    & (64.91, 64.91) & (56.17, 56.79) \\
No Algorithm & (56.79, 56.17)   & (---, ---) \\
\bottomrule
\end{tabular}
\end{minipage}

\bigskip

\begin{minipage}{0.5\textwidth}
\centering
\begin{tabular}{lcc}
\multicolumn{3}{c}{\sc Recommendation (SG5)} \\
\toprule
 & Algorithm & No Algorithm \\
\midrule
Algorithm    & (66.18, 66.18) & (68.95, 66.67) \\
No Algorithm & (66.67, 68.95)   & (65.71, 65.71) \\
\bottomrule
\end{tabular}
\end{minipage}
\end{table}

\begin{table}[H]
\centering
\caption{Punishment durations in AH/AA markets}\label{tab:punishment_durations_AA_AH}
\centering
\begin{tabular}[t]{llrrrrrr}
\toprule
Treatment & Market Type & n & mean & median & sd & min & max\\
\midrule
Outsourcing & AH & 618 & 1.25 & 1 & 0.71 & 1 & 7\\
Recommendation & AA & 1022 & 1.68 & 1 & 2.08 & 1 & 39\\
Recommendation & AH & 203 & 3.06 & 2 & 3.30 & 1 & 25\\
\bottomrule
\end{tabular}
\end{table}

\begin{table}[H]
\centering
\caption{First punishment phase duration in \textsc{Recommendation} by supergame}
\label{tab:first_punishment_phase}

\begin{tabular}{rrrrrrr}
\toprule
SG & n & mean & median & sd & min & max\\
\midrule
All & 234 & 2.01 & 1 & 2.00 & 1 & 21\\
1 & 70 & 2.00 & 1 & 2.57 & 1 & 21\\
2 & 45 & 1.82 & 1 & 1.19 & 1 & 6\\
3 & 15 & 1.00 & 1 & 0.00 & 1 & 1\\
4 & 48 & 2.44 & 2 & 2.31 & 1 & 10\\
\addlinespace
5 & 56 & 2.09 & 2 & 1.59 & 1 & 9\\
\bottomrule
\end{tabular}
\end{table}

\begin{table}[H]
\centering
\caption{Adherence with first recommendation after punishment phase}
\label{tab:compliance_after_punishment}
\begin{threeparttable}

\begin{tabular}[t]{lrrr}
\toprule
SG & Adherence & n & Share\\
\midrule
All & No & 147 & 0.63\\
 & Yes & 87 & 0.37\\
\addlinespace[-0.5em]\\
1 & No & 44 & 0.63\\
 & Yes & 26 & 0.37\\
\addlinespace[-0.5em]\\
2 & No & 26 & 0.58\\
 & Yes & 19 & 0.42\\
\addlinespace[-0.5em]\\
3 & No & 12 & 0.80\\
 & Yes & 3 & 0.20\\
\addlinespace[-0.5em]\\
4 & No & 32 & 0.67\\
 & Yes & 16 & 0.33\\
\addlinespace[-0.5em]\\
5 & No & 33 & 0.59\\
 & Yes & 23 & 0.41\\
\bottomrule
\end{tabular}
\begin{tablenotes}[para, flushleft]
\footnotesize Note: Adherence indicates whether participants followed the first algorithmic recommendation to cooperate after exiting the punishment phase in the \textsc{Recommendation} treatment.
\end{tablenotes}
\end{threeparttable}
\end{table}

\begin{table}[H]
\caption{Market prices in SG1: \textsc{Baseline} vs. \textsc{Outsourcing}-HH markets}
\begin{center}
\begin{tabular}{l c}
\toprule
 & Market Price \\
\midrule
Outsourcing-HH          & $-0.88^{**}$ \\
                        & $(0.40)$     \\
(Intercept)             & $2.52^{***}$ \\
                        & $(0.12)$     \\
\midrule
Sub-sample & Base \& Out-HH (SG1) \\
$\# $ observations & 3024 \\
$\# $ cluster & 14 \\
R$^2$ & 0.03 \\
\bottomrule
\multicolumn{2}{p{0.5\textwidth}}{\scriptsize{Results are based on linear regression with standard errors clustered at the matching-group level. We use a sub-sample of Baseline and Outsourcing-HH in Supergame 1 to assess the market price difference. Significance levels: *** \( p < 0.01 \); ** \( p < 0.05 \); * \( p < 0.1 \).}}
\end{tabular}\label{tab:market_prices_sg1_baseline_outHH}
\end{center}
\end{table}

\end{document}